\documentclass[showpacs,prl,onecolumn,aps,superscriptaddress,preprintnumbers,nofootinbib]{revtex4}
\usepackage[T1]{fontenc}
\usepackage[latin9]{inputenc}
\usepackage{amsmath,amssymb,bigints}
\usepackage{epsfig}
\usepackage{graphicx}
\usepackage{amsmath}
\usepackage{amsfonts}
\def\slashchar#1{\setbox0=\hbox{$#1$}     		
   \dimen0=\wd0                                 	
   \setbox1=\hbox{/} \dimen1=\wd1               	
   \ifdim\dimen0>\dimen1                        	
      \rlap{\hbox to \dimen0{\hfil/\hfil}}      	
      #1                                        	
   \else                                        	
      \rlap{\hbox to \dimen1{\hfil$#1$\hfil}}   	
      /                                         	
   \fi}

\newcommand{\bra}[1]{\left \langle #1 \right |}
\newcommand{\ket}[1]{\left | #1 \right \rangle}

\renewcommand{\vec}{\boldsymbol}
\newcommand{\beq}{\begin{equation}}
\newcommand{\eeq}{\end{equation}}
\newcommand{\bea}{\begin{eqnarray}}
\newcommand{\eea}{\end{eqnarray}}
\newcommand{\baa}{\begin{array}}
\newcommand{\eaa}{\end{array}}
\renewcommand{\vec}[1]{\boldsymbol{#1}}
\def\eq#1{{Eq.~(\ref{#1})}}
\def\fig#1{{Fig.~\ref{#1}}}

\newcommand{\bas}{\bar{\alpha}_S}
\newcommand{\as}{\alpha_S}

\newcommand{\nn}{\nonumber}

\newcommand{\h}{\frac{1}{2}}

\newcommand{\Lb}{\left(}
\newcommand{\Rb}{\right)}
\def\pom{{I\!\!P}}

\newcommand{\T} {\mbox{T}}
\renewcommand{\vec}[1]{\boldsymbol{#1}}

\def\pom{{I\!\!P}}

\def\VEV#1{\left\langle #1\right\rangle}

\begin{document}

\title{Large  impact parameter behaviour  in the CGC/saturation 
approach: 
 a new non-linear equation.}
\author{E. ~Gotsman}
\email{gotsman@post.tau.ac.il}
\affiliation{Department of Particle Physics, School of Physics and Astronomy,
Raymond and Beverly Sackler
 Faculty of Exact Science, Tel Aviv University, Tel Aviv, 69978, Israel}
 \author{E.~ Levin}
\email{leving@tauex.tau.ac.il, eugeny.levin@usm.cl}
\affiliation{Department of Particle Physics, School of Physics and Astronomy,
Raymond and Beverly Sackler
 Faculty of Exact Science, Tel Aviv University, Tel Aviv, 69978, Israel}\affiliation{ Departamento de F\'\i sica,
Universidad T$\acute{e}$cnica Federico Santa Mar\'\i a   and
Centro Cient\'\i fico-Tecnol$\acute{o}$gico de Valpara\'\i so,
Casilla 110-V,  Valparaiso, Chile}

\date{\today}


\begin{abstract}
 In this paper we propose a solution to the long standing problem  in 
 the CGC/saturation approach: the power-like fall off of the scattering
 amplitudes at large $b$. 
 We propose a new non-linear equation, which takes into account
 random walks both in transverse momenta of the produced gluons and in
  their impact parameters.
 We demonstrate, that this equation is in accord with previous attempts
 to include the diffusion  in impact parameters in the BFKL evolution equation.
  We show in the paper, that the solution to a new equation 
  results in the exponential    decrease  of the scattering amplitude
 at large  impact parameter, and in the restoration of the Froissart 
theorem.  
 \end{abstract}
\preprint{TAUP - 3040/19}
\maketitle


\tableofcontents
\section{ Introduction}
It is well known that perturbative QCD   has a
  fundamental problem:
 the scattering amplitude decreases at large impact parameters ($b$) 
as a
 power of $b$.  In particular,  the CGC/saturation approach\cite{KOLEB}, 
 which is based on perturbative QCD,  is confronted by this problem. 
 At large $b$ the scattering amplitude is small and, therefore,  only
 the linear BFKL equation\cite{BFKL} is determined by  the scattering 
amplitude
 in perturbative QCD.  It is known that the eigenfunction of this equation
 (the scattering amplitude of two dipoles with sizes $r$ and $R$) has the
 following form\cite{LIP}
\beq \label{EIGENF}
\phi_\gamma\Lb \vec{r} , \vec{R}, \vec{b}\Rb\,\,\,=\,\,\,\Lb \frac{
 r^2\,R^2}{\Lb \vec{b}  + \h(\vec{r} - \vec{R})\Rb^2\,\Lb \vec{b} 
 -  \h(\vec{r} - \vec{R})\Rb^2}\Rb^\gamma\,\,\xrightarrow{b\,\gg\,
r,R}\,\,\Lb \frac{ r^2\,R^2}{b^4}\Rb^\gamma\,\,\equiv\,\,e^{\gamma\,\xi}
~~\mbox{with}~~ \xi\,=\,\ln \Lb \frac{ r^2\,R^2}{b^4}\Rb
\eeq
One can see that at large impact parameter $b$ the amplitude has a
 power-like decrease,  which leads to the violation of the Froissart
 theorem\cite{FROI}.   The violation of the Froissart theorem stems
 from the growth of the radius of interaction as a power of the
 energy. Since in Ref.\cite{LIP}  it was proven that  the eigenfunction
 of any kernel  with conformal symmetry  has the form of \eq{EIGENF},
 we can only change the large $b$ behaviour  by
 introducing a new dimensional scale  in the kernel of the equation. 
The problem has been known from the beginning of the saturation
 approach\cite{LERYB1,LERYB2}  and several ideas have been proposed, of 
 how to introduce a new dimensional scale in the kernel of the BFKL
 equation (See Refs.\cite{LERYB1,LERYB2,LETAN,QCD2,KHLE,KKL}). However,
  for the high energy community at large,  the problem  was 
appreciated 
 only after  the papers of Refs.\cite{KW,FIIM} were published,   where 
 it
 was demonstrated,  that the violation of the Froissart theorem cannot
 be avoided in the framework of the CGC approach.

 First, we wish to illustrate why the Froissart theorem is violated 
for the BFKL equation.  The general solution to the BFKL  for the
 dipole scattering amplitude equation, has the following form:
 
 \beq \label{BFKL}
 N\Lb  r ,  Y; b \Rb\,\,\,=\,\,\,\int^{\epsilon + i \infty}_{\epsilon
 - i \infty}\frac{d \gamma}{2\,\pi\,i} e^{\omega\Lb \bas,\gamma\Rb \,Y
 } \,\phi_\gamma\Lb \vec{r} , \vec{R}, \vec{b}\Rb\phi_{\rm in}\Lb \gamma\Rb
\,\,=\,\,\int^{\epsilon + i \infty}_{\epsilon - i \infty}\frac{d
 \gamma}{2\,\pi\,i} e^{\omega\Lb \bas,\gamma\Rb \,Y \,\,+\,\,\gamma\,\xi} 
\,\phi_{\rm in}\Lb \gamma\Rb \eeq 
 where
 \bea \label{CHI}
\omega\Lb \bas, \gamma\Rb\,\,&=&\,\,\bas\,\chi\Lb \gamma \Rb\,\,\,=\,\,\,\bas \Lb 2 \psi\Lb 1\Rb \,-\,\psi\Lb \gamma\Rb\,-\,\psi\Lb 1 - \gamma\Rb\Rb\\
&\,\xrightarrow{\gamma \to \h}& \,\omega_0\,\,+\,\,D\,\Lb \gamma - \h\Rb^2  \,\,+\,\,{\cal O}\Lb (\gamma - \h)^3\Rb\,\,=\,\,\bas 4 \ln 2  \,\,+\,\,\bas 14 \zeta\Lb 3\Rb \Lb \gamma - \h\Rb^2  \,\,+\,\,{\cal O}\Lb (\gamma - \h)^3\Rb\nn
 \eea
 $\psi(z)$  denotes the Euler psi-function
 $\psi\Lb z\Rb = d \ln \Gamma(z)/d z$.
 
The main contribution in \eq{BFKL} stems from $\gamma \to \h$,
 where we can use the expansion shown in \eq{CHI}
 evaluating this integral using the method of steepest descend one 
can see
 that saddle point occurs at $\gamma_{\rm SP} = \h -
 \frac{\xi}{2 D Y}\,\,\ll\,\,1$ and the amplitude is equal to
 \beq \label{FR1}
  N\Lb  r ,  Y; b \Rb\,\,\,\xrightarrow{b\,\gg\,r,R}\,\,\,\,\,\phi\Lb\h\Rb\Lb \frac{ r^2\,R^2}{b^4}\Rb^{1/2}\frac{1}{ 2 \sqrt{D Y}}e^{\omega_0 Y\,-\,\frac{\xi^2}{4 D Y}}
  \eeq

Using \eq{FR1} we can attempt  to  determine the upper bound from the
 unitarity constraints
\beq \label{UNIT}
2 \,N\Lb  r,  Y; b \Rb\,\,=\,\,|N\Lb  r ,  Y; b \Rb|^2 \,\,+\,\,G_{in} \Lb 
 r^2 ,  Y; b \Rb
\eeq
where $G_{in}$ describes the contribution of all inelastic
 processes. Recalling that $N$ is the imaginary part of the
 scattering amplitude and assuming that the real part of the
 amplitude is small, as is  the case for the BFKL equation,
 one can see that the unitarity constraint has the solution:
\beq \label{FR2}
N = 1 - \exp\Lb - \Omega\Lb r,Y;  b\Rb\Rb\,\,<\,\,1; ~~~~~~~~~G_{in} \,=\,1
 \,-\,\exp\Lb -2 \Omega\Lb r,Y;  b\Rb \Rb
\eeq
where $\Omega > 0$ denotes an arbitrary function.

Now we can find the bound for the total cross section following 
Ref.\cite{FROI}.

\beq \label{FR3}
\sigma_{tot}\,\,=\,\,2 \int N\Lb r,Y; b\Rb d^2 b\,\,<\,\,2 \int^{b_0} d^2 b\,\,+\,\,\int_{b_0} d^2 b N\Lb r,Y; b\Rb 
\eeq
We need to solve the following equation to find the value of $b_0$, 

\beq \label{FR4}
 N\Lb  r ,  Y; b_0 \Rb\,=\,\,\phi\Lb\h\Rb\Lb \frac{ r^2\,R^2}{b^4}\Rb^{1/2}\frac{1}{ 2 \sqrt{D Y}}e^{\omega_0 Y\,-\,\frac{\xi^2}{4 D Y}}\,=\,f_0\,<1
 \eeq

At large $Y$ this solution gives $b^2_0 \propto e^{ \Lb D + \sqrt{D (D +
 4 \omega_0)}\Rb Y}$ and therefore,
\beq \label{FR5}
 N\Lb  r ,  Y; b_0 \Rb\,\,<\,\,e^{ \Lb D + \sqrt{D (D + 4 \omega_0)}\Rb Y}
 \eeq
 
 Note, that if for the soft amplitude of the typical form 
 $N_{\rm soft} \propto e^{\omega_0 Y - \mu b}$  and  $b_0 \,\propto Y$.
 One can see, that  the first integral in \eq{FR3} leads to $\sigma\, 
\leq\, {\rm Const}\,\, y^2$ , which is the Froissart theorem.
 This amplitude violates the Froissart theorem  and can be considered
 only at large values of $b$ where  $N_{\rm soft} < 1$. However, the 
eikonal
 solution of \eq{FR2}  with $\Omega = N_{soft}$,  satisfies the unitarity
 constraints and leads to  the amplitude, that describes the Froissart
 disc: the amplitude is equal to 1 for $b  < (1/\mu)Y$ and it has an
 edge which behaves as $N _{\rm soft} $.

 We hope that this  simple estimate shows,  that the power-like
 decrease of the scattering amplitude is the source of the problem.
  To solve the problem in the framework of the CGC/ saturation approach we 
need to introduce a new 
dimensional scale $\mu$, 
 in addition to the saturation momentum. 
 In Ref.\cite{FROI}
 it is shown that this new scale is related to the mass of the
 lightest hadrons. Perturbative QCD cannot 
reproduce
 the observed spectrum of hadrons,  and attempting to solve this 
problem, we 
are doomed to introduce
 something from non-perturbative QCD estimates.
  Since the non-perturbative approach is still in  an 
 embryonic stage,  one can only guess how to introduce this 
scale, which depends
  crucially  on non-perturbative estimates in
 lattice QCD, on phenomenology of high energy interactions and on
 intuition, which comes from considering different theoretical models.
 We hope that more or less full list of the attempts to solve this
 problem, and to find and introduce a new dimensional scale appear
 in Refs.\cite{LERYB1,LERYB2,LETAN,QCD2,KHLE,KKL,FIIM,GBS1,BLT,GKLMN,HAMU,
 MUMU,BEST1,BEST2,KOLE,LETA,LLS,LEPION,KAN}.
 
 At first sight the recent papers\cite{CCM,BCCM} have questioned the
 need of  a new dimensional scale, since they demonstrate that the
 next-to-leading order BFKL equation generates the  exponential type
 of the impact parameter behavior  without the need of  a new 
dimensional scale. However, it turns out\cite{CLM}  that
 the NLO corrections do not change 
the power-like decrease of the scattering amplitude  at large impact 
parameter, generating the exponential-type decreases in the large but
 limited  region of the values of the impact parameter.\footnote{In
 addition to the power-like behaviour at large $b$ the NLO corrections
lead to an oscillating behaviour of the scattering amplitude at
 large $b$, in  direct contradiction   with   the unitarity 
constraint\cite{CLM}.}

    In this paper we re-visit  one of the possible ways  of 
introducing
 a new dimensional scale: to incorporate in the BFKL equation the
 diffusion in  impact parameters($b$).  The first such attempt was
 undertaken in the distant 1990's \cite{LERYB1,LERYB2} and during three
 decades  we have worked in this area
    \cite{LETAN,QCD2,KHLE,KKL,BLT,LEPION,KAN}.  This is the reason 
that in
 the first part of the paper we give a brief review  of these efforts,
 while in the second part we propose our generalization of the BFKL 
 equation, which takes into account the diffusion on $b$ in accord
 with QCD estimates. We will discuss the structure of the scattering
 amplitude at high energies, which at the moment   appears to be a 
black disc
 with the radius increasing as a power of energy.  This paper is partly
 motivated by Ref.\cite{KAN}, in which many questions  concerning QCD 
has been
 formulated  on the black disc behaviour at high energy\footnote{ We will
 use  the Froissart disc instead the black disc behaviour with  the radius
 which increase as $\ln(s)$.}
    from the point of  boost invariance and the parton model. We hope
 that we answer some of these questions in this paper. In particular, we 
 will demonstrate  that QCD leads to the Froissart disc at high energies,
 with specific behaviour of the amplitude at the edge of this disc.


\section{Diffusions}


\subsection{Regge approach and Gribov's diffusion in impact parameters}
In the framework of the  Regge approach the high energy amplitude is given
 by the exchange of the Pomeron, and has the following
 form\cite{COL,SOFT,GRLEC,LEPOM}:
\beq \label{GD1}
s\,N\Lb s,Q_t\Rb \equiv\,\,{\rm Im} A\Lb s, Q_T\Rb\,\,=\,\,g_1\Lb Q_T\Rb \,g_2\Lb Q_T\Rb e^{ \alpha_\pom\Lb Q^2_T\Rb Y}
\eeq
where $g_1, g_2$ and trajectory
 $\alpha_\pom\Lb Q^2_T\Rb \equiv\,\,1\,+\,\Delta_{\pom}\Lb Q_T\Rb 
 = 1 + \Delta_\pom - \alpha'_\pom Q^2_T  + {\cal O}\Lb  Q^4_T \Rb$
 are the functions that have to be taken from the phenomenology.
 $Y = \ln\Lb s\Rb$  and $Q_T$ is the momentum transferred by the
 Pomeron \footnote{In the case of the deep inelastic processes $Y
 = \ln\Lb 1/x\Rb$, where $x$ is the Bjorken variable.}. \eq{GD1}
 can be viewed as the solution to the following equation:
\beq \label{GD2}
\frac{d N\Lb Y,Q_T\Rb}{ d Y}\,\,=\,\,\Delta_\pom\Lb Q_T\Rb\,N \Lb Y,Q_T\Rb
\eeq
with the initial condition 
\beq \label{GD3}
N \Lb Y=0,Q_T\Rb\,\,=\,\,g_1\Lb Q_T\Rb \,g_2\Lb Q_T\Rb
\eeq
In the impact parameter representation solution of \eq{GD1} takes the form:
\beq\label{GD4}
 N\Lb Y, b\Rb\,
\,=\,\,\int \frac{d^2 Q_T}{\Lb 2 \pi\Rb^2}\,e^{ i \vec{b} \cdot
 \vec{Q}_T}\,N \Lb Y=0,Q_T\Rb\,\,=\,\,g_1 g_2 e^{\Delta_\pom,Y}
  n\Lb Y,b \Rb
~~~~\mbox{with}~~~
n\Lb Y, b\Rb\,\,=\,\,  \frac{1}{4 \pi \alpha'_\pom Y} e^{ -
 \frac{b^2}{4 \alpha'_{\pom} Y}}
\eeq
 In \eq{GD4} we have neglected the $Q_T$ dependence of $g_1$ and $g_2$
 which does not contribute at high energies.
 
\begin{figure}[ht]
     \begin{center}
     \includegraphics[width=0.5\textwidth]{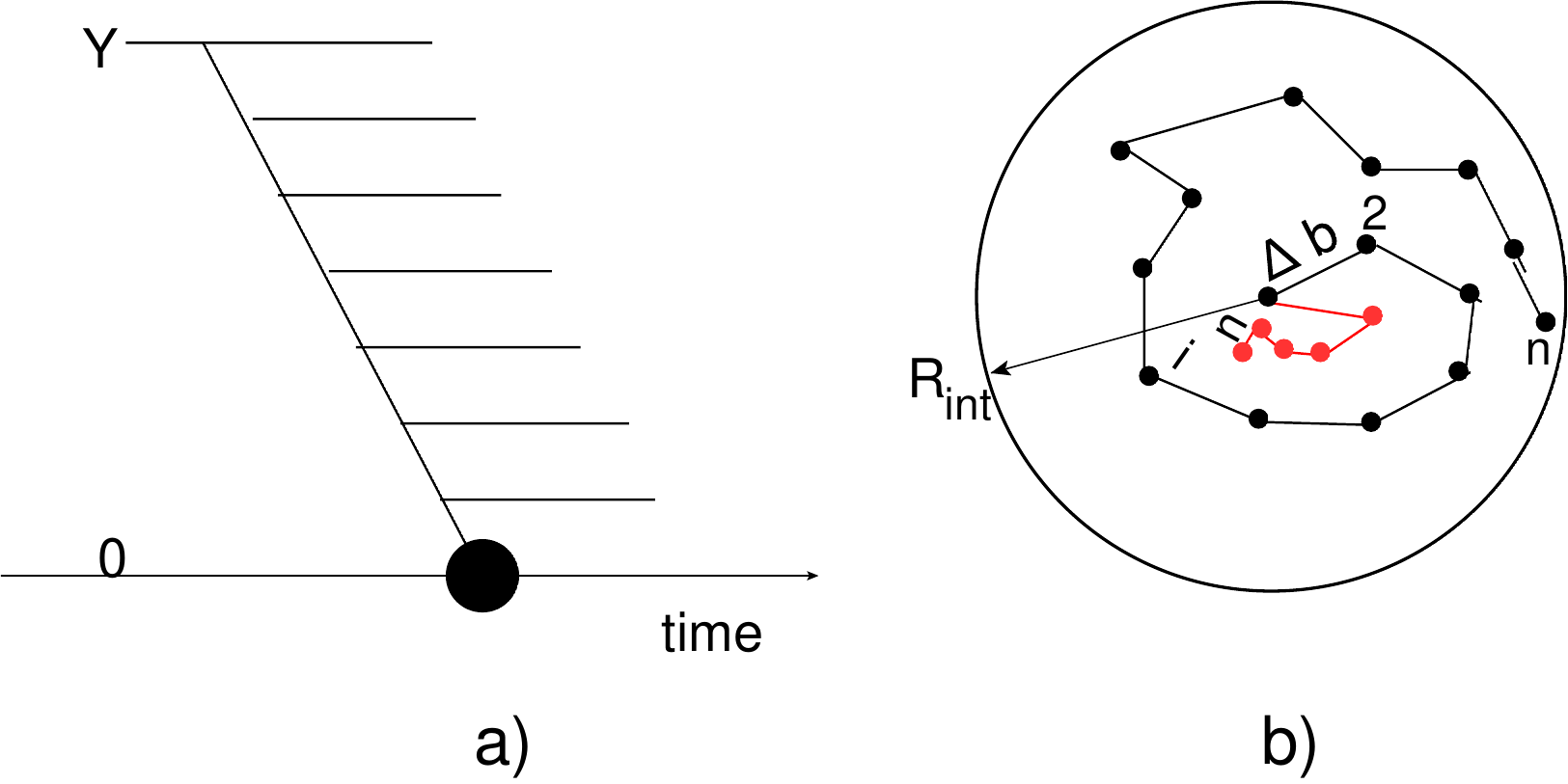} 
     \end{center}    
      \caption{ The structure of the parton cascade:  \fig{gdif}-a shows
 the time structure of the cascade while \fig{gdif}-b
      illustrates the random walk in $b$.}
\label{gdif}
   \end{figure}

 In Ref.\cite{GRLEC}  the simple fact is noted i.e.  $n\Lb Y, b\Rb$
 is the solution of the diffusion equation:
 \beq \label{GD5}
 \frac{ d n\Lb Y, b\Rb }{d Y}\,\,=\,\,\alpha'_\pom \nabla^2_b\, n\Lb Y, b\Rb
 \eeq
 \eq{GD5} together with \eq{GD1} for the total cross section:
 \beq \label{GD6}
 \sigma_{tot} = 2\,g_1 \,g_2\, N\Lb Y,Q_T=0\Rb\,=\,2 \,\,
\,g_1\,g_2\,e^{ \Delta_\pom Y}\,\,=\,\,\sigma_0\,\sum_{n=0}^\infty
 \frac{ \Delta_\pom Y}{n!} \,\,
 \eeq
 have very simple interpretations in the parton model. 
In the parton model\cite{GRLEC,LEPOM,FEYN} it is  assumed,
 that we can describe the interaction by a field theory,
 in which all integrals over transverse momenta are convergent,
 and they lead to the mean transverse momentum, which does not
 depend on energy. In such a theory,  the contribution to the
 total cross section of the scattering amplitude for production
 of $n$ partons in each order of perturbation approach,  can be
 viewed as
 \beq \label{GD7}
 \sigma_n\,\,=\,\,M_{2 \to 2+n}\Lb \{p_{i,T}\}\Rb \prod^n_{i=0} 
 d^2 p_{i,T} d y_i\,\,=\,\,M_{2 \to 2+n}\Lb \{p_{i,T}\}\Rb \prod^n_{i=0}
  d^2 p_{i,T}\,\frac{Y^n}{n!}
 \eeq
 In \eq{GD7} we assume that in the proposed theory  the amplitude  is not
 equal to zero,  when rapidities of emitted partons are equal to zero,
 and choose the largest contribution which comes from the ordering
 \beq \label{GD8}
 0 <   y_1 <  y_2  <  \dots  < y_i < \dots  < y_n < Y
 \eeq
 One can see that in \eq{GD6} $ M_{2 \to 2+n}\Lb \{p_{i,T}\}\Rb
 \prod^n_{i=0}  d^2 p_{i,T}\,=\,\Delta_\pom^n$
 \footnote{  In this estimate we assume that
  $ \sigma_0 $ in \eq{GD4} does not depend on energy.
 In the field theories which can be a realization of
 the parton model, usually $\sigma_0 \propto 1/s^2$ but
  the first result from QCD was the understanding that
 in this approach $\sigma_0$ does not depend on energy\cite{LONU}.}
 and from 
 this equation we can conclude that the number of emitted
 partons  $n \,=\,\Delta_\pom \,Y$. The Gribov idea that
 the emission of partons  has no other correlations
 except the fixed transverse momentum,  can be viewed
 as a random walk in two dimensional space. For each emission
 due to uncertainty principle 
 \beq \label{GD9}
 \Delta b_i p_{i,T} \sim 1 ~~~~\mbox{or}~~~ \Delta b_i \sim
 \frac{1}{<|p_{i,T}|> }
 \eeq
 Therefore, after each emission the position of the parton will be
 shifted by an amount $\Delta b^2$ from \eq{GD9},
which is  on average the same. After $n$ emissions, we have the
 picture shown in \fig{gdif}, and the total shift in $b$ is equal to
\beq \label{GD10}
R^2_{int} \,\,=\,\,b^2_n \,\,=\,\,\frac{1}{<|p_{i,T}^2|> }\,\,=
\,\,\frac{1}{<|p_{i,T}^2|> } \Delta_\pom Y
\eeq
Therefore, this diffusion reproduces the shrinkage of the
 diffraction peak. Indeed, 
\beq \label{GD11}
R^2 \,\,=\,\,\frac{\int d^2 b \,b^2\, N\Lb Y,b\Rb}{\int d^2
 b \, N\Lb Y,b\Rb}\,\,=\,\,4\,\alpha'_\pom\,Y
\eeq
Comparing \eq{GD10} and \eq{GD11} one can see that 
\beq \label{GD12}
\alpha'_\pom \,=\,\,\frac{\Delta_\pom}{4 <|p_{i,T}^2|>}
\eeq

\eq{GD2} can be re-written in the impact parameter representation
 for $N\Lb Y, b\Rb$ of \eq{GD4}
\beq \label{GD13}
\frac{ d  N\Lb Y, b\Rb}{d Y}\,\,=\,\,\int d^2 b' \,K\Lb \vec{b}
 \,-\,\vec{b}'\Rb\,N\Lb Y, b'\Rb ~~~~~\mbox{where}\,\,K\Lb b\Rb \,\,=
\,\int \frac{d^2 Q_T}{(2 \pi)^2} e^{i  \vec{b}\, \cdot \,\vec{Q}_T } 
\Delta_\pom\Lb Q_T\Rb
\eeq

Let us write the equation for the radius of interaction  (see \eq{GD11}) .
 First, we  see that for $ \int d^2 b \,b^2\,N\Lb Y, b\Rb =  R^2 \int
 d^2 b N\Lb Y, b \Rb$ we have the following equation:
\bea \label{GD14}
&&\frac{ d \Lb  R^2 \int d^2 b N\Lb Y, b \Rb\Rb}{ d Y} \,\,=
\,\,\int d^2 b\, d^2 b' \, b^2 \,K\Lb \vec{b} \,-\,\vec{b}'\Rb\,N\Lb Y, b'
 \Rb\nn\\
&& \,=\,\int d^2(\vec{b} - \vec{b}')\, d^2 b'\,K\Lb \vec{b} \,-\,\vec{b}'\Rb \Bigg( ((\vec{b} - \vec{b}')^2 \,+ 2 (\vec{b} - \vec{b}')\cdot \vec{b}' \,+\,b'^2\Bigg)N\Lb Y, b' \Rb\,\,\nn\\
&&=\,\,\underbrace{\Lb \int d^2(\vec{b} - \vec{b}')\, \,K\Lb \vec{b} \,-\,\vec{b}'\Rb  ((\vec{b} - \vec{b}')^2\Rb}_{ <| \Delta b|> \Delta_\pom\Lb Q_T=0\Rb}  \int d^2 b' N\Lb Y, b' \Rb\,\,+\,\,\underbrace{\int d^2(\vec{b} - \vec{b}')\, d^2 b'\,K\Lb \vec{b} \,-\,\vec{b}'\Rb (\vec{b} - \vec{b}')\cdot \vec{b}'  N\Lb Y, b' \Rb}_{\mbox{= 0}}\nn\\
&&\,+\,\,\underbrace{\Lb \int d^2(\vec{b} - \vec{b}')\, K\Lb \vec{b} \,-\,\vec{b}'\Rb \Rb}_{\Delta\Lb Q_T=0\Rb}\Lb \int d^2 b' \, b'^2 \,N\Lb Y, b' \Rb\Rb
\eea

The second term vanishes due to integration over the angle and finally
 we have the following equation

\beq \label{GD15}
\frac{ d R^2}{d Y} \,\,=\,\, \Delta_{\pom} <| \Delta b^2|> 
\eeq
with the solution $R^2 \,=\, \Delta_{\pom} <| \Delta b^2|> Y\,\,=
\,\,-\nabla^2_{Q_T} \Delta_\pom\Lb Q_T\Rb|_{Q_T =0}\,Y$.

\begin{figure}[ht]
     \begin{center}
     \includegraphics[width=0.6\textwidth]{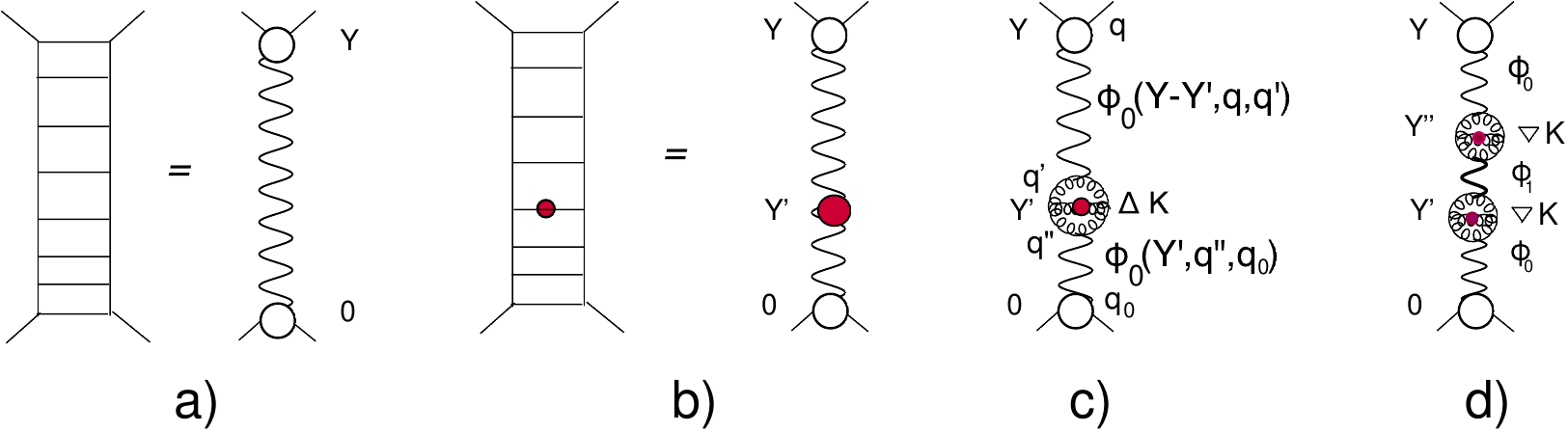} 
     \end{center}    
      \caption{ The Mueller diagrams for calculation of
 $\nabla^2_{Q_T} N\Lb Y,Q_T\Rb|_{Q_T=0}$. The wavy lines denote
 the Pomeron contribution at $Q_T = 0$. For \fig{mudi}-a and 
\fig{mudi}-b  it is the Pomeron in the parton model of \eq{GD1},
 while for \fig{mudi}-c and \fig{mudi}-d it is  the BFKL Pomeron
 in QCD(see \eq{S1}. The red blobs denote $\nabla^2_{Q_T} 
\Delta_\pom\Lb Q_T\Rb|_{Q_T= 0}$ or $\nabla_{Q_T} \Delta_\pom\Lb
 Q_T\Rb|_{Q_T= 0}$ as it is indicated in the figures. Functions
 $\phi_0$ and $\phi_1$ are defined in the text.}
\label{mudi}
   \end{figure}

We can obtain \eq{GD15}  using the Mueller diagrams\cite{MUDI} of 
\fig{mudi}.
 
 Indeed, one can see that 
 \beq \label{GD16}
 \nabla^2_{Q_T} N\Lb Y,Q_T\Rb|_{Q_T=0}\,\,=
\,\,-\int d Y' e^{\Delta_\pom (Y - Y')}\,e^{\Delta_\pom  Y')}\nabla^2_{Q_T}
 \Delta_\pom\Lb Q_T\Rb|_{Q_T=0}\,\,=\,\,  4 \alpha'_\pom Y  
 e^{\Delta_\pom Y} 
 \eeq
 Therefore, for $R^2$ we obtain $R^2 =  4 \alpha'_\pom\, Y 
  e^{\Delta_\pom Y} /e^{\Delta_\pom Y} =  \,4 \alpha'_\pom Y$ 
in accord with \eq{GD15}.

 The amplitude of \eq{GD4}   increases with energy and violates the
 unitary constraints (see \eq{UNIT}).
 The eikonal  unitarization of \eq{FR2} leads to the following
 amplitude
 \beq \label{GD17}
 N\Lb Y, b\Rb\,\,=\,\,1\,\,-\,\,\exp\Lb - e^{\Delta_\pom\,Y}
 \, \frac{1}{4 \pi \alpha'_\pom Y} e^{ - \frac{b^2}{4 \alpha'_{\pom}
 Y}}\Rb
 \eeq
 
 One can see that this amplitude tends to unity at $b >\,\,\,\,2
\,\sqrt{\Delta_\pom\,\alpha'_\pom}\,Y $ leading to the total cross
 section $\sigma_{tot} \,\propto\,Y^2$ in accord with the Froissart
 theorem\cite{FROI}.
 
 In spite of the primitive level of calculations, especially if you
 compare them with typical QCD calculations in DIS, the parton model
 was a good guide for the Pomeron structure for years and, 
  it is still the model where we can see all typical features
 of  the soft  Pomeron.  It turns out that the time structure of the
 parton cascade is preserved for QCD  and  simple parton estimates 
can help
 develop our intuition  regarding  the solution of the QCD problems.
 \begin{boldmath}
\subsection{BFKL approach  and  diffusion in  $\ln\Lb p_T\Rb$.}
\end{boldmath}

\subsubsection{The BFKL equation.}

The BFKL equation was derived in momentum representation\cite{BFKL}
 and has the following form:
\beq \label{BFKLM}
\frac{\partial \widetilde{N}\Lb Y; q, Q_T\Rb}{ \partial Y}\,\,=\,\,\bas
 \Bigg(\int \frac{d^2 q'}{2 \pi}  K_{\rm em}\Lb \vec{q} - \vec{q}',
 \vec{Q}_T\Rb \widetilde{N}\Lb Y; q', Q_T\Rb\,\,-\,\,K_{\rm reg}\Lb \vec{q}
 - \vec{q}', \vec{Q}_T\Rb \widetilde{N}\Lb Y; q', Q_T\Rb\Bigg)
\eeq
where $\bas = \Lb N_c/\pi\Rb \as$.  Kernel $K_{\rm em}$ describes
 the emission of a gluon, while  kernel $K_{\rm reg}$ is responsible
 for the reggeization of gluons in t-channel.
They have the forms:
\bea \label{KERM}
K_{\rm em}\Lb \vec{q} - \vec{q}', \vec{Q}_T\Rb\,\,&=&\,\,\h \frac{1}{\Lb \vec{q} - \vec{q}'\Rb^2} \Bigg\{ - \frac{Q^2_T 
\Lb \vec{q} - \vec{q}'\Rb^2 }{\Lb \vec{Q}_T - \vec{q}'\Rb^2\,q'^2}  \,+1\,+ \frac{\Lb \vec{Q}_T - \vec{q}\Rb^2\,q'^2}{\Lb \vec{Q}_T - \vec{q}'\Rb^2\,q^2}\Bigg\} \xrightarrow{Q_T =0} \frac{1}{\Lb \vec{q} - \vec{q}'\Rb^2} \\
K_{\rm reg}\Lb \vec{q} - \vec{q}', \vec{Q}_T\Rb\,\,&=&\h \frac{1}{\Lb \vec{q} - \vec{q}'\Rb^2} \Bigg\{  \frac{q^2 
}{\Lb \vec{q} - \vec{q}'\Rb^2\,+\,q'^2}\,\,+\,\, \frac{\Lb \vec{Q}_T \,-\,\vec{q}\Rb^2 
}{\Lb \vec{q} - \vec{q}'\Rb^2\,+\,\Lb \vec{Q}_T - \vec{q}'\Rb^2}\Bigg\} \xrightarrow{Q_T =0}\frac{1}{\Lb \vec{q} - \vec{q}'\Rb^2}   \frac{q^2 
}{\Lb \vec{q} - \vec{q}'\Rb^2\,+\,q'^2}\nn
\eea

This equation is rewritten in the coordinate representation for
 the scattering amplitude  of a dipole with the size $r$ at
 impact parameter $\vec{b}$\cite{LIP,MUCD}:
\beq \label{BFKLC}
\frac{\partial}{\partial Y} N\Lb \vec{r}, \vec{b} , 
 Y \Rb =\bas\!\! \int \frac{d^2 \vec{r}'}{2\,\pi}\,K\Lb \vec{r}',
 \vec{r} - \vec{r}'; \vec{r}\Rb \Bigg(N\Lb \vec{r}',\vec{b} - \h
 \Lb \vec{r} - \vec{r}' \Rb, Y\Rb + 
N\Lb\vec{r} - \vec{r}', \vec{b} - \h \vec{r}', Y\Rb \,\,- \,\,N\Lb \vec{r},\vec{b},Y \Rb\Bigg)
\eeq
with
\beq \label{KERC}
K\Lb \vec{r}', \vec{r} - \vec{r}'; \vec{r}\Rb \,\,=\,\,\frac{r^2}{r'^2\,\Lb \vec{r} - \vec{r}'\Rb^2}
\eeq

In \eq{BFKLC} 
\beq \label{MC}
N\Lb \vec{r}, \vec{b} ,  Y \Rb\,\,=\,\,r^2\,\int \frac{d^2 q}{2\,\pi} e^{i \vec{q} \cdot \vec{r}}\widetilde{N}\Lb q, b, Y\Rb
;
~~~~~~ \widetilde{N}\Lb Y; q, Q_T\Rb\,\,=\,\,\int \frac{d^2 b}{(2\,\pi)^2} e^{i \vec{Q}_T \cdot \vec{b}}\widetilde{N}\Lb q, b, Y\Rb;
\eeq

 \begin{boldmath}
\subsubsection{Solutions and  random walk in  $\ln\Lb p_T\Rb$.}
\end{boldmath}

We have discussed solutions in the  coordinate representation 
 (see
 \eq{EIGENF},  \eq{BFKL} and \eq{FR1}). These solutions are very
 useful for discussions of the non-linear corrections, since the
 unitarity constraints are diagonal for the dipole scattering
 amplitude (see \eq{UNIT}). However, discussing the random walk
 in $\ln\Lb p_T\Rb$ we need a solution in the momentum representation
 in which  $p_{i,T}$ are the momenta of produced gluons. Note, that in 
\eq{KERM}
 $\vec{q} - \vec{q}' = \vec{p}_T$. For $Q_T$ the solution can be
   easily obtained, since the eigenfunction have the following 
form\cite{BFKL}
\beq \label{EIGENFM}
\phi_\gamma\Lb q, q_0, Q_T=0\Rb\,\,=\,\,\Lb\frac{q^2}{q^2_0}\Rb^{{\gamma} - 1}
\eeq
Comparing \eq{EIGENFM} with \eq{EIGENF} one can see that
 $q_0 \sim 1/R$, where $R$ denotes the size of the target. Repeating all 
the
 steps, that are given by \eq{BFKL} and \eq{CHI},  we obtain the
 solution in the form of \eq{FR1}; viz.
 \beq \label{S1}
  \widetilde{N}\Lb  q ,  Q_T=0, Y  \Rb\,\,\,=\,\,\,\phi\Lb\h\Rb\Lb
 \frac{ q^2_0}{q^2}\Rb^{1/2}\,e^{\omega_0 \,Y}\widetilde{n}\Lb q,
 Y\Rb
  ~~~\mbox{where}~~~\widetilde{n}\Lb q, Y\Rb\,\,=\,\,  
  \frac{1}{ 2 \sqrt{\pi \,D\, Y}} \,e^{-\,\frac{\tilde{\xi}^2}{4 D Y}}
  \eeq
with $\tilde{\xi} = \ln\Lb q^2/q^2_0\Rb$. One can recognize that for the
 function $n$ we have the diffusion equation in the form:
\beq \label{S2}
\frac{\partial}{\partial Y} \widetilde{n}\Lb \tilde{\xi} ,Y\Rb\,\,=\,\,D \frac{\partial^2}{\partial \tilde{\xi}^2} \widetilde{n}\Lb \tilde{\xi} ,Y\Rb
\eeq
Therefore, the BFKL equation describes that at each emission,
 $\ln\Lb q^2/q'^2\Rb$  changes its value by a constant $(\ln\Lb
 q^2/q'^2\Rb)^2 =\delta$, and as a result after $n$ emissions we
 obtain $\ln^2\Lb p^2_T/q^2_0\Rb\,=\,\delta \,n$. Since $n =
 \omega_0 \,Y$ (see the previous section) we obtain $(\ln\Lb
 q^2/q^2_0\Rb)^2\,\,\delta \omega_0 \,Y $. This estimate shows
 that $4 D = \delta \omega_0$. From this estimate we see that
 after $n$ emissions the typical transverse momenta increase
 as $< |p^2_T|>\,\,=q^2_0 \exp\Lb \delta \,n\Rb$, making the
 shift in $b$\, $<|\Delta b^2|>_n \propto 1/< |p^2_T|> \sim
 (1/q^2_0)  \exp\Lb - \delta n\Rb$. Therefore, only  a small
 number of steps at the beginning could participate in the
 increase of $b$ (see red lines in \fig{gdif}-b ).

~

\subsubsection{The Green function   of the BFKL Pomeron.}


The solution of \eq{S1} can be re-written in the following form:
\beq\label{GF1}
  \widetilde{N}\Lb  q ,  Q_T=0, Y  \Rb\,\,\,=\,\,\,\int^{\epsilon + i \infty}_{\epsilon - i \infty} \frac{d \omega}{ 2 \pi i}
\,\int^{\epsilon + i \infty}_{\epsilon - i \infty} \frac{d \gamma}{ 2 \pi i} \frac{1}{ \omega - \omega\Lb \gamma\Rb}e^{ \omega Y  + \Lb \gamma - 1\Rb \tilde{\xi}} \phi_{int}\Lb \gamma\Rb
\eeq
We introduce the Green function of the BFKL Pomeron as follows:
\beq\label{GF1}
G^{\rm BFKL}\Lb \omega, \tilde{\xi}\Rb \,\,\,=\,\,\,
\,\int^{\epsilon + i \infty}_{\epsilon - i \infty} \frac{d \gamma}{ 2 \pi i} \frac{1}{ \omega - \omega\Lb \gamma\Rb}e^{ \omega Y  + \Lb \gamma - 1\Rb \tilde{\xi}}
\eeq
 The Green function in $Y$ representation can be calculated as
\bea \label{GF2}
\hspace{-0,5cm}G^{\rm BFKL}\Lb Y, \tilde{\xi} \Rb \,\,\,&=&\,\,\,\frac{1}{q^2_0}\int^{\epsilon + i \infty}_{\epsilon - i \infty} \frac{d \omega}{ 2 \pi i}
G^{\rm BFKL}\Lb \omega, \tilde{\xi};Q_T=0\Rb\,\,
=\,\,\,\frac{1}{q^2_0}\int^{\epsilon + i \infty}_{\epsilon - i \infty} \frac{d \omega}{ 2 \pi i}\,\int^{\epsilon + i \infty}_{\epsilon - i \infty} \frac{d \gamma}{ 2 \pi i} \frac{1}{ \omega - \omega\Lb \gamma\Rb}e^{ \omega Y  + \Lb \gamma - 1\Rb \tilde{\xi}}\nn\\
 &\xrightarrow{\gamma = \h + i \nu,\nu \ll 1}& \frac{1}{q\,q_0} e^{\omega_0\,Y}  \int^{\epsilon + i \infty}_{\epsilon - i \infty}\frac{d \omega}{ 2\,\pi\,i}\,\frac{1}{2\,D \,\kappa_0}  e^{ - D \,\kappa^2_0\, Y\,+\,\kappa_0\,\,\tilde{\xi}}~~~\mbox{where}~~~\kappa_0\,\,=\,\,\sqrt{\frac{\Lb \omega\,\,-\,\,\omega_0\Rb}{ D}}\eea
Integrating over $\kappa_0 $ we  obtain the solution of \eq{S1}.

One can see that   $G^{\rm BFKL}\Lb Y=0, \tilde{\xi};Q_T=0\Rb
 =\delta\Lb \tilde{\xi}\Rb$  and therefore, the scattering amplitude
 can be found $\widetilde{N}\Lb Y; q, Q_T=0 \Rb\,\,=\,\,\int  d
 \tilde{\xi}_0 \,G^{\rm BFKL}\Lb \tilde{\xi} - \tilde{\xi}_0\Rb\,\Phi_{in}\Lb \tilde{\xi}_0\Rb$, where $\Phi$ is  the initial condition for the scattering amplitude.
It should  also be mentioned that  factors $1/(q \,q_0)$ are
 absorbed in
 integration of $\tilde{\xi}$ in the  diagrams
for Pomeron interactions.
 \begin{boldmath}
\subsubsection{The BFKL approach: random walk in  $b $.}
\end{boldmath}

 As one can see from \eq{S1} we  have introduced a new
 dimensional scale: $q_0$.  It was introduced as the non-perturbative
 size of the target $q_0 \sim 1/R$, but its actual meaning is  the
 separation scale: in perturbative QCD we can calculate only for
 $p_T> q_0$, while smaller transverse momenta have to be treated
 in  non-perturbative approaches. From the solution of \eq{S1} one
 can see that the probabilty to have $p_T = q_0$ is small  $
 \propto\,\frac{1}{2 \sqrt{D\,Y}}$
but not negligible. The gluons with transverse  momenta
 of the order of $q_0$ could participate in the random
 walk in $b$, leading to \cite{LERYB1,LERYB2}
\beq \label{RWB1}
<| b^2 |>_n\,=\,\frac{1}{q^2_0} P^n_{q_0} \,n\,\,\propto\,\,\,\frac{1}{q^2_0}  \,\frac{1}{\sqrt{n}} \,n \propto \frac{1}{q^2_0} \sqrt{Y}
\eeq
where $P^n_{q_0}$ denotes the probability to have the minimum momentum
 after $n$ emissions.
In Refs.\cite{LERYB1,LERYB2} the numerical coefficient in \eq{RWB1}
 is evaluated. Since the average $<| b^2 |>_n$  diverges at small $q$,
 in some sense the value of 
the coefficient is not important.
  However,  we feel it is instructive
 to understand two qualitative features:  the infrared divergency
 and the energy dependence.

First, we calculate the main ingredients of the calculation that we
 have discussed in section II-A :
\bea \label{DELTAK}
\nabla_{Q_T}K_{\rm em}\Lb \vec{q} - \vec{q}', \vec{Q}_T\Rb\Big{|}_{Q_T=0}
 \,&=&-\frac{1}{\Lb \vec{q} - \vec{q}'\Rb^2}\Bigg( \frac{\vec{q}}{q^2} \,-\,\frac{\vec{q'}}{q'^2}\Bigg);\nn\\
\nabla_{Q_T}K_{\rm reg}\Lb \vec{q} - \vec{q}', \vec{Q}_T\Rb\Big{|}_{Q_t=0} \,&=&-\frac{1}{\Lb \vec{q} - \vec{q}'\Rb^2}\Bigg( \frac{\vec{q}}{q'^2} \,-\,\frac{\vec{q'}\,q^2}{q'^4}\Bigg);\nn\\
-\nabla^2_{Q_T}K_{\rm em}\Lb \vec{q} - \vec{q}', \vec{Q}_T\Rb\Big{|}_{Q_T=0} \,&=&\,-\Delta_{Q_T}K_{\rm em}\Lb \vec{q} - \vec{q}', \vec{Q}_T\Rb\Big{|}_{Q_T=0}\,\,=\,\,\frac{2}{q^2\,q'^2};\nn\\
\Delta_{Q_T} K_{\rm reg}\Lb \vec{q} - \vec{q}', \vec{Q}_T\Rb\Big{|}_{Q_T=0}\,\,&=&\,\,0
\eea

As we have discussed in section II-A (see \eq{GD14}),  we can write the
 equation for
 $-\nabla^2_{Q_T}\widetilde{N}\Lb Y; q, Q_T\Rb|_{Q_T = 0} \,\,=\,\,\int
 \frac{d^2 b}{(2\,\pi)^2}\, b^2\,\widetilde{N}\Lb q, b, Y\Rb$ applying
 $ -\nabla^2_{Q_T}$ to the both parts of \eq{BFKLM}.
 In doing so, we obtain:
 \bea \label{RWB2}
 \frac{ \partial }{ \partial Y} \Lb -\nabla^2_{Q_T}\widetilde{N}\Lb Y;
 q, Q_T\Rb|_{Q_T = 0}\Rb &=& \int d^2 q' \nabla^2_{Q_T}K\Lb \vec{q} -
 \vec{q}', \vec{Q}_T\Rb\Big{|}_{Q_T=0}\widetilde{N}\Lb Y; q', Q_T\Rb|_{Q_T = 0} \nn\\
 &-& 2  \int d^2 q' \nabla_{Q_T}K\Lb \vec{q} - \vec{q}',
 \vec{Q}_T\Rb\Big{|}_{Q_T=0}\cdot\Lb \nabla_{Q_T}\widetilde{N}\Lb Y;
 q', Q_T\Rb|_{Q_T = 0}\Rb\nn\\ 
 &\,\,+\,\,&
 \int d^2 q' K\Lb \vec{q} - \vec{q}', \vec{Q}_T\Rb\Big{|}_{Q_T=0}\Lb 
 -\nabla^2_{Q_T}\widetilde{N}\Lb Y; q', Q_T\Rb|_{Q_T = 0}\Rb 
 \eea
 To obtain the complete system of equations we need to add the equation 
for  
$\nabla_{Q_T}\widetilde{N}\Lb Y; q',
 Q_T\Rb|_{Q_T = 0}$.
 It takes the form:
  \bea \label{RWB3}
 \frac{ \partial }{ \partial Y} \Lb \nabla_{Q_T}\widetilde{N}\Lb Y; q, Q_T\Rb|_{Q_T = 0}\Rb &=& \int d^2 q' \nabla_{Q_T}K\Lb \vec{q} - \vec{q}', \vec{Q}_T\Rb\Big{|}_{Q_T=0}\widetilde{N}\Lb Y; q', Q_T\Rb|_{Q_T = 0} \nn\\
 &+&   \int d^2 q' K\Lb \vec{q} - \vec{q}', \vec{Q}_T\Rb\Big{|}_{Q_T=0}\Lb \nabla_{Q_T}\widetilde{N}\Lb Y; q', Q_T\Rb|_{Q_T = 0}\Rb 
 \eea 
  \eq{RWB3} can be re-written as
  \beq \label{RWB4}
  \Lb \frac{\partial }{\partial Y} - 1\Rb \nabla_{Q_T}\widetilde{N}\Lb Y;  q, Q_T\Rb|_{Q_T = 0}  \,=\,\bas\frac{\vec{q}}{q^2}\Bigg(\int_{q^2} \frac {d q'^2}{q'^2}\widetilde{N}\Lb Y; q', Q_T=0\Rb  \,-\,\widetilde{N}\Lb Y; q, Q_T=0\Rb \Bigg)
  \eeq
  In  the double Mellin transform of \eq{GF1}  the solution to \eq{RWB4} 
has the
 form:
  \beq \label{RWB5}
  \nabla_{Q_T}\widetilde{N}\Lb \omega;  \gamma, Q_T\Rb|_{Q_T = 0} \,=\,\bas\,\frac{\vec{q}}{q^2}\frac{1}{\omega - \omega\Lb \gamma\Rb} \Bigg( \frac{1}{1 - \gamma}  \,-\,1\Bigg)
  \eeq
  
  Plugging the solution of \eq{RWB5} into \eq{RWB2}, we reduce this equation
 to the following one:
  \bea \label{RWB6}
&&   \frac{ \partial }{ \partial Y} \Lb -\nabla^2_{Q_T}\widetilde{N}\Lb Y; q, Q_T\Rb|_{Q_T = 0}\Rb\,\,-\,\, \int d^2 q' K\Lb \vec{q} - \vec{q}', \vec{Q}_T=0\Rb\Lb  -\nabla^2_{Q_T}\widetilde{N}\Lb Y; q', Q_T\Rb|_{Q_T = 0}\Rb =\nn\\
   &&\,\,2\,\frac{\bas}{q^2}\Bigg\{ \int^\infty_{0}   \frac{d q'^2}{q'^2} \widetilde{N}\Lb Y; q', Q_T=0\Rb  \,+\,2\,\bas
  \int^{q^2}_{0}   \frac{d q'^2}{q'^2}  \int^{\epsilon + i \infty}_{\epsilon - i \infty} \frac{d \gamma}{ 2 \pi i} \frac{1}{ \omega - \omega\Lb \gamma\Rb}\Bigg( \frac{1}{1 - \gamma}  \,-\,1\Bigg)
e^{ \omega Y  + \Lb \gamma - 1\Rb \tilde{\xi}} \Bigg\} 
\eea
The main contributions to the integrals  over $q'$ stems from the region
 $ q' \to 0$, Taking this into account the solution in  the $\omega$-
 representation takes the form:
  \beq \label{RWB7}
\nabla^2_{Q_T}\widetilde{N}\Lb \omega ; q, Q_T\Rb|_{Q_T = 0}\,\,=\,\,
2\,\frac{\bas}{q^2} \frac{1 + \bas}{ D^2 \kappa_0^2}\,\,\propto\,
\,\frac{1}{\omega - \omega_0}
\eeq
Therefore, we see from \eq{RWB7} that  $\nabla^2_{Q_T}\widetilde{N}\Lb
 \omega ; q, Q_T\Rb|_{Q_T = 0} \propto e^{\omega_0 Y} $

We need to divide this solution  by the $\widetilde{N}\Lb Y; q',
 Q_T\Rb|_{Q_T = 0}$
and finally, we obtain  for $\xi^2 \ll D \,Y$ that
\beq \label{RWB8} 
\VEV{b^2}\,\,=\,\,2  \bas(1  + \bas) \,\sqrt{\frac{\pi}{D}}\,\frac{1}{q^2}
\,\sqrt{Y} \,\,=\,\,\alpha'_{\rm eff}\,\sqrt{Y}\eeq

As was shown in Refs\cite{LERYB1,LERYB2},  the second term does not
 have the suppression of the order of $\bas$, as we can see from 
\eq{RWB7}. 
 Such an enhancement comes from the replacement  in \eq{RWB2} 
\bea \label{RWB9}
&&2  \int d^2 q' \nabla_{Q_T}K\Lb \vec{q} - \vec{q}', \vec{Q}_T\Rb\Big{|}_{Q_T=0}\cdot\Lb \nabla_{Q_T}\widetilde{N}\Lb Y; q', Q_T\Rb|_{Q_T = 0}\Rb\,\to\,\nn\\
&&2  \int d^2 q' \nabla_{Q_T}K\Lb \vec{q} - \vec{q}', \vec{Q}_T\Rb\Big{|}_{Q_T=0}\bigotimes  \phi_1\Lb  Y - Y',q",q',\theta\Rb \bigotimes \Lb \nabla_{Q_T}\widetilde{N}\Lb Y'; q", Q_T\Rb|_{Q_T = 0}\Rb
\eea
$\bigotimes$ stands for all needed integrations. The eigenfunction
 $\phi_1$\cite{LIP} depends on the angles between $\vec{q}"$ and
 $\vec{q}'$ and has an intercept which is negative and $\propto
 \,\bas$. Integration over $Y'$ leads to the factor $1/\bas$ which
 compensates the small  value of $\bas$.
This contribution actually leads to the change of the numerical
 coefficient,  which does not have  much meaning, since $\VEV{b^2}$
 is infrared unstable and $q \to q_0$ gives the main contribution.
The value of the scale $q_0$ is not determined in our approach. 
 Hence, we introduce a new dimensional scale $\alpha'_{\rm eff}$
 which, we hope, will heal the large $b$ behaviour of the scattering
 amplitude.

~


\section{Searching for a new dimensional scale  in  the 
CGC/saturation 
approach}

In this section we would like to give a brief review of the attempts to
 introduce both the random walk in $\ln\Lb p_T\Rb$, and in $b$, in the
 effective theory describing QCD at high energies: the CGC/saturation
 approach. These efforts cover a long span in time, from the first
 GLR equation\cite{GLR}(see also Refs.\cite{MUQI,MV,MUCD})  till 
 recent papers.  Our goal is to introduce the random walk in $b$ in
 the framework of the non-linear evolution equations, which guarantee
 the unitarity of the scattering amplitude.

\subsection{Non-linear evolution equations in QCD}
We  start with the current form of the non-linear equation\cite{BK},
 which has the following form for the scattering amplitude of the
 dipole with size $\vec{r}$  and rapidity $Y$ at  impact parameter
 $\vec{b}$:

\bea \label{BKC}
\frac{\partial}{\partial Y} N\Lb \vec{r}, \vec{b} ,  Y \Rb &=&\bas\!\! \int \frac{d^2 \vec{r}'}{2\,\pi}\,K\Lb \vec{r}', \vec{r} - \vec{r}'; \vec{r}\Rb \Bigg\{N\Lb \vec{r}',\vec{b} - \h \Lb \vec{r} - \vec{r}' \Rb, Y\Rb + 
N\Lb\vec{r} - \vec{r}', \vec{b} - \h \vec{r}', Y\Rb \,\,- \,\,N\Lb \vec{r},\vec{b},Y \Rb\nn\\
& &~~~~~~~~~~~~~~~~~~~~~~~~~~~~~~~~~ - N\Lb\vec{r} - \vec{r}', \vec{b} - \h \vec{r}', Y\Rb   \, N\Lb \vec{r}',\vec{b} - \h \Lb \vec{r} - \vec{r}' \Rb, Y\Rb\Bigg\}\eea
The kernel  $K\Lb \vec{r}', \vec{r} - \vec{r}'; \vec{r}\Rb$ is given by
 \eq{KERC}.
For $\widetilde{N}\Lb Y; q, Q_T\Rb $ and $\widetilde{N}\Lb q, b, Y\Rb$ (see \eq{MC} ) can be re-written in the forms:
\bea \label{BKMM}
\frac{\partial \widetilde{N}\Lb Y; q, Q_T\Rb}{ \partial Y}\,\,&=&\,\,\bas \Bigg\{\int \frac{d^2 q'}{2 \pi}  K_{\rm em}\Lb \vec{q} - \vec{q}', \vec{Q}_T\Rb \widetilde{N}\Lb Y; q', Q_T\Rb\,\,-\,\,K_{\rm reg}\Lb \vec{q} - \vec{q}', \vec{Q}_T\Rb \widetilde{N}\Lb Y; q', Q_T\Rb\nn\\
  &-& \,\,\int \frac{d^2 Q'_T}{4 \pi^2} \widetilde{N}\Lb Y; q', \vec{Q}_T - \vec{Q}'_T\Rb \,\widetilde{N}\Lb Y; q', \vec{Q}'_T\Rb\Bigg\}
  \eea
with the kernels from \eq{KERM}.

Finally, for $b \,\gg\,max\{ r, r'\} $ \eq{BKMM} can be simplified
 (see Ref.\cite{KOLEB}) to
\bea  \label{BKM}
\frac{\partial \widetilde{N}\Lb Y; q, b\Rb}{ \partial Y} &=& \\
& &  \bas \Bigg\{\int \frac{d^2 q'}{2 \pi}  K_{\rm em}\Lb \vec{q} - \vec{q}',Q_T=0\Rb \widetilde{N}\Lb Y; q', b\Rb\,\,-\,\,K_{\rm reg}\Lb \vec{q} - \vec{q}',Q_T=0\Rb \widetilde{N}\Lb Y; q, b\Rb \,\,-\,\,
\Lb \widetilde{N}\Lb Y; q', b \Rb \Rb\Bigg\}\nn
\eea

\subsection{ The GLR equation}

\begin{figure}[ht]
     \begin{center}
     \includegraphics[width=0.2\textwidth]{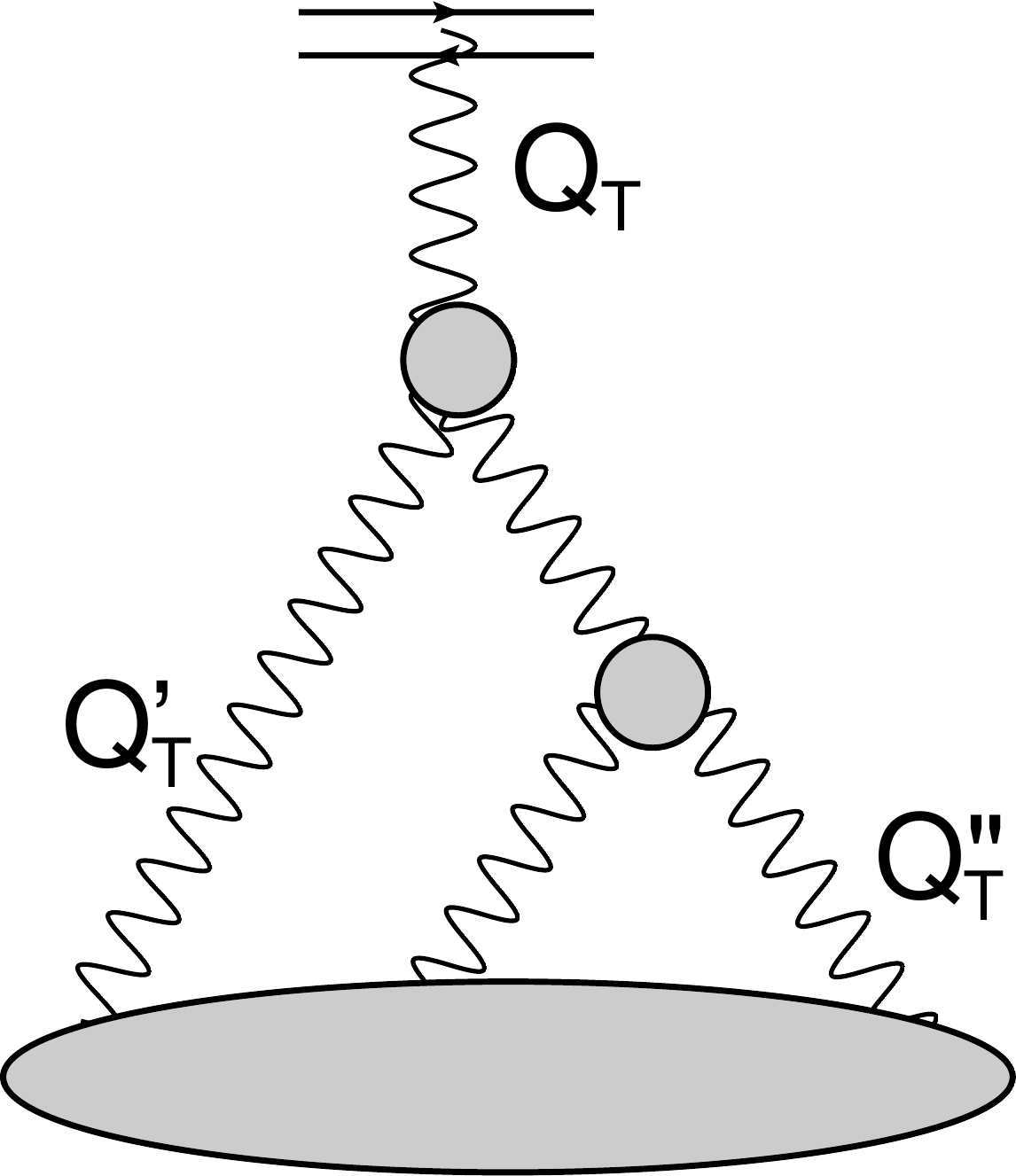} 
     \end{center}    
      \caption{An  example of  `fan' diagrams, whose sum is described
 by the non-linear equations. The wavy lines denote the BFKL Pomerons. }
\label{fan}
   \end{figure}

The first non-linear equation \cite{GLR}  addressed the problem
 of large $b$ behaviour of the scattering amplitude.  The non-linear
 evolution equations of the previous subsections  can be viewed as the
 sum of the `fan' diagrams (see \fig{fan} for the example of such diagrams).
 It was shown in Ref.\cite{GLR} that integrations over all transfer
  momenta $Q'_T$ carried by the BFKL Pomerons  in such diagrams,
 (see \fig{fan} for example) turns out to be of the order of
 $ Q"_T \sim 1/R_h$, where $R_h$ is the size of the hadron target.
 Therefore, if the dipole sizes are much smaller than $R_h$, as it is
 for  deep inelastic processes, we can replace $
 \widetilde{N}\Lb Y; q, Q_T\Rb$ in \eq{BKMM} by
 $ \widetilde{N}\Lb Y; q, Q_T=0\Rb S\Lb Q_T\Rb $
 and reduce \eq{BKMM} to the GLR equation
\bea \label{GLR}
\hspace{-0.4cm}&&\frac{\partial \widetilde{N}\Lb Y; q, Q_T=0\Rb}{ \partial Y}\,=\\
\hspace{-0.4cm}&&\,\,\bas \Bigg\{\int \frac{d^2 q'}{2 \pi}  K_{\rm em}\Lb \vec{q} - \vec{q}', \vec{Q}_T=0\Rb \widetilde{N}\Lb Y; q', Q_T\Rb\,\,-\,\,K_{\rm reg}\Lb \vec{q} - \vec{q}', \vec{Q}_T=0\Rb \widetilde{N}\Lb Y; q', Q_T=0\Rb
  - \,\,\frac{1}{S_h}  \widetilde{N}^2\Lb Y; q', \vec{Q}_T =0\Rb \Bigg\}\nn
  \eea
where $ \frac{1}{S_h} \,=\,\int \frac{d^2 Q'_T}{4 \pi^2} S^2\Lb Q'_T\Rb$ is
 a constant which depends on the unknown non-perturbative structure of the
 hadrons. $S_h $ is the area of the hadron which is populated by gluons.

  The GLR equation is written for the infrared safe observable: the
 gluon structure function;  and it  was suggested that it replace the 
DGLAP
 equation in the region of small $x$( high energies). Considering this
 equation we do not have a problem of the large $b$ behaviour of the
 amplitude. The intuition behind this equation is very transparent:
 in deep inelastic processes the number of gluons increases with the
 growth of energy, while the size of their distribution remains the
 same until the hadron 
 becomes black.  At ultra high energies the radius of the distribution
 could increase and we cannot trust the equation. From \eq{GLR} we
 can see that the large $b$ problem is 
  intimately  
   related to the black disc behaviour of the scattering amplitude,
 and we need to develop a different equation to describe the
 behaviour of the edge of the disc.
   
\subsection{The first equation incorporating both diffusions}
  In Ref.\cite{LERYB1}  the first equation was  proposed that
 incorporates both diffusions in $\ln\Lb p_T\Rb$ and in $b$.
 This equation has  the form:
  \beq \label{FEQ1}
\frac{\partial \widetilde{N}\Lb Y; q, b\Rb}{ \partial Y} \,\,=\,\,  \bas \Bigg\{\int \frac{d^2 q'}{2 \pi} \int \frac{d^2 b'}{2 \pi} K\Lb \vec{q} - \vec{q}'\Rb\Lb  q'^2 \,\kappa \,e^{-  \kappa  \,q'^2\, (\vec{b} - \vec{b}')^2} \Rb
\widetilde{N}\Lb Y; q', b'\Rb\,\,\,-\,\,
 \widetilde{N}^2\Lb Y; q, b \Rb\Bigg\}\nn
\eeq  
  where 
  \beq \label{K}
  K\Lb \vec{q} - \vec{q}'\Rb\,\,=\,\,\frac{1}{\Lb \vec{q} - \vec{q}'\Rb^2}\,\,-\,\,\frac{1}{\Lb \vec{q} - \vec{q};\Rb^2\,\,\Lb\Lb \vec{q} - \vec{q}'\Rb^2  + q'^2\Rb}  \delta^{(2)}\Lb \vec{q} - \vec{q}' \Rb  
  \eeq

  At large $b$ the shadowing corrections (the non-linear term) in
 the equation makes a  small contribution, and one can see that the new
 kernel leads to the same intercept of the BFKL Pomeron since it does
 not contribute to  $\int d^2 \widetilde{N}\Lb Y; q, b\Rb \equiv 
 \widetilde{N}\Lb Y; q, Q_T=0\Rb $. 
  
  Recently, the differential form of \eq{FEQ1} has been proposed  in
 Ref.\cite{KAN}, which  has  the form  \footnote{We take the liberty of 
presenting
 this equation in a  slightly  different form than in Ref.\cite{KAN}. 
This form
 coincides with the original one in the diffusion approximation for the
 BFKL kernel.} :
   \beq \label{FEQ2}
\frac{\partial \widetilde{N}\Lb Y; q, b\Rb}{ \partial Y} \,\,=\,\,  \bas \Bigg\{\int \frac{d^2 q'}{2 \pi} \int \frac{d^2 b'}{2 \pi} K\Lb \vec{q} - \vec{q}'\Rb\widetilde{N}\Lb Y; q', b'\Rb\,\,\,-\,\,
 \widetilde{N}^2\Lb Y; q, b \Rb\Bigg\}  \,+\, \frac{\kappa}{q^2}\,\nabla^2_b \widetilde{N}\Lb Y; q, b\Rb\eeq   
  
 Neglecting the non-linear term at large $b$ and using the function
 $\widetilde{N}^2\Lb \omega; \gamma , Q_T \Rb$ we can re-write \eq{FEQ2}
 in the form:
 \beq \label{FEQ3}
\Lb  \omega - \omega(\gamma) \,-\, \alpha'_{\rm eff}\, Q^2_T\Rb\widetilde{N}\Lb \omega; \gamma , Q_T \Rb = 0
\eeq  
with $\alpha'_{\rm eff} = \kappa/q^2_0$. In \eq{FEQ3} we replace $1/q^2 $
 by $1/q^2_0$ where $q_0$ is the smallest momentum where one can use the
 perturbative QCD approach.

  Hence the solution takes the form:
  \bea \label{FEQ4}
   \widetilde{N}\Lb Y; q, b \Rb\,\,&=&\,\,  \int^{\epsilon + i \infty}_{\epsilon - i \infty}\frac{d \gamma}{2 \pi \,i}\int\frac{d^2Q_T}{4 \pi^2}  e^{ \omega(\gamma) \,Y\, -\,(1 - \gamma) \xi \,+\,i
    \vec{Q}_T\cdot \vec{b} \,-\,\alpha'_{\rm eff} Q^2_T }\widetilde{n}_{in}\Lb  \gamma \Rb\nn\\
    &=& \frac{q_0}{q}  \widetilde{n}_{in}\Lb \h\Rb\,\Lb\frac{1}{2 \sqrt{ \pi\,D\,Y}} e^{\omega_0 \,Y \,-\,\frac{\xi^2}{4\,D\,Y} }\Rb\,\Lb \frac{1}{4 \pi \alpha'_{\rm eff}\,Y} e^{ - \frac{b^2}{4\,\alpha'_{\rm eff}\,Y}}\Rb
    \eea
  Therefore, we see that indeed we have both diffusions, but the diffusion
 in $b$ does not have $\VEV{b^2} \propto \sqrt{Y}$. \eq{FEQ4} leads to the
 solution with a black disc, whose radius can be found from the 
equation:
  \beq \label{FEQ5}
  \omega_0 \,Y \,-\,\frac{\xi^2}{4\,D\,Y}  - \frac{b^2_0}{4 \,\alpha'_{\rm eff} Y}\,\,=\,\,0;~~~~  
 ~~~ b^2_0 \,=\, 4 \,\alpha'_{\rm eff} \, \omega_0 \,Y^2 \,\,- \,\, \frac{\alpha'_{\rm eff} }{D} \,\xi^2;
   \eeq

    As we have demonstrated in section II-A  \eq{FEQ5} results in the
 Froissart bound;
      \beq \label{FEQ5}
      \sigma \,\,\leq\,\,2 \pi b^2_0 = 2 \pi \Lb    4 \,\alpha'_{\rm eff} \, \omega_0 \,Y^2 \,\,- \,\, \frac{\alpha'_{\rm eff} }{D} \,\xi^2\Rb
      \eeq

  The main problem with  \eq{FEQ1} and \eq{FEQ2} is, that they do not
 reproduce the correct $\VEV{b^2} \, \propto\, \sqrt{Y}$  which
 as we have discussed,  arises in QCD.
  
  At first sight \eq{FEQ1} can lead to a different solution, but 
 for $ \widetilde{N}^2\Lb \omega; \gamma , Q_T \Rb$  it takes the
 following form:
   \beq \label{FEQ6}
\Lb  \omega - \omega(\gamma) \,+\,e^{- \alpha'_{\rm eff}\, Q^2_T}-1\Rb\widetilde{N}\Lb \omega; \gamma , Q_T \Rb = 0
\eeq    
  The solution to \eq{FEQ6} is
    \bea \label{FEQ7}
   \widetilde{N}\Lb Y; q, b \Rb\,\,&=&\,\,  \int^{\epsilon + i \infty}_{\epsilon - i \infty}\frac{d \gamma}{2 \pi \,i}\int\frac{d^2Q_T}{4 \pi^2}  e^{ \omega(\gamma) \,Y\, -\,(1 - \gamma) \xi \,+\,i
    \vec{Q}_T\cdot \vec{b} \,+\,\Lb e^{-\,\alpha'_{\rm eff} Q^2_T }-1\Rb\,Y}\widetilde{n}_{in}\Lb  \gamma \Rb\nn\\   
  &=& \frac{q_0}{q}  \widetilde{n}_{in}\Lb \h\Rb\,\Lb\frac{1}{2 \sqrt{ \pi\,D\,Y}} e^{\omega_0 \,Y \,-\,\frac{\xi^2}{4\,D\,Y} }\Rb\,\sum^\infty_{n=0}\Lb \frac{n}{4 \pi \alpha'_{\rm eff}\,Y} \frac{Y^n}{n!}\,e^{ - \frac{b^2}{4\,\alpha'_{\rm eff}\,n}}\Rb\nn\\
  &=& \frac{q_0}{q}  \widetilde{n}_{in}\Lb \h\Rb\,\Lb\frac{1}{2 \sqrt{ \pi\,D\,Y}} e^{\omega_0 \,Y \,-\,\frac{\xi^2}{4\,D\,Y} }\Rb\,\Lb \frac{1}{4 \pi \alpha'_{\rm eff}\,Y} e^{ - \frac{b^2}{4\,\alpha'_{\rm eff}\,Y}}\Rb    \eea  
  
  In \eq{FEQ7} we summed over $n$ using the method of steepest
 descend and used ,
  that at $Q_T =0$, the $\omega(\gamma)$ spectrum is the same as
 in the BFKL equation.

\subsection{ The field theory realization of the parton model: the
 BFKL Pomeron in  spontaneously broken QCD in 2+1 space-time dimensions }
  In Ref.\cite{QCD2} the BFKL Pomeron is studied in spontaneously
 broken 2 + 1 QCD, The coupling constant of this theory $g$ has
 dimension of mass, therefore, this is a super renormalizable theory
 which is, actually, a theoretical realization of the parton model, that
 has been discussed in section II-A. In this theory we do not have the
 ultraviolet  divergencies, which results in the absence of the scaling
 violations, in the behaviour of the deep inelastic structure functions.
 On the other hand, the infrared singularities in the massless limit
 are stronger than in 3+1 QCD. Being such, this theory is suitable  for
 discussing the influence of the infrared singularities on the Pomeron
 structure. In particular, we expect  Gribov's diffusion in the
 impact parameters.
  
  The BFKL equation in this theory has the same form of \eq{BFKLM} with
 replacement
  \beq \label{QCD21}
  \frac{d^2 q'}{4 \pi^2} \,\,\to\,\,  \frac{d q'}{2 \pi}  
  \eeq
  and with the change of the kernels of \eq{KERM} due to inclusion of the
 mass of gluon $m$. It can be written as
  \bea \label{QCD22}
 && \Lb \omega - \alpha\Lb q\Rb - \alpha\Lb Q_T -  q\Rb\Rb  \widetilde{N}\Lb Y; q, Q_T - q\Rb\,\,=\nn\\
 &&\,\,\frac{\omega}{A\Lb Q_T\Rb} \,+\,\frac{g^2 N_c}{ 8 \pi^2} \int \frac{d\, q'}{\Lb q'^2 + m^2  \Rb^2 \,\Lb \Lb q - q' \Rb^2 + m^2\Rb} K\Lb q, q',Q_T\Rb \widetilde{N}\Lb Y; q', Q_T - q'\Rb  
 \eea
 where the gluon trajectories are
 \beq \label{QCD23}
\alpha\Lb q \Rb\,=\,-\frac{g^2 N_c}{ 8 \pi^2}\Lb q^2 + m^2\Rb\int \frac{ d\, q'}{\Lb q'^2 + m^2 \Rb\,\Lb ( q - q')^2 + m^2\Rb}\,=\,\as \frac{q^2 + m^2}{q^2 + 4 m^2}
 \eeq
 where $\as = g^2\,N_c/(4 \pi m)$ ($N_c $ denotes the number of colours).
\beq \label{QCD24}
K\Lb q, q - q', Q_T\Rb\,\,=\,\,A\Lb Q_T\Rb + \frac{2}{ (q - q')^2 +
 m^2}\Bigg( \frac{q^2 + m^2}{q'^2 + m^2} + \frac{(Q_T -q)^2 + m^2}{Q_T 
-q')^2 + m^2}\Bigg)~~\mbox{with} ~~A(Q_T)= - 2\Lb Q^2_T + \frac{N_c+1}{N_c} \,m^2\Rb
\eeq

  \eq{QCD22} is solved  analytically in Ref.\cite{QCD2},  both in
 coordinate and momentum representations. It turns out that the
 rightmost singularity in $\omega$ is the Regge pole with the
 trajectory
  \beq \label{QCD25}
  \alpha_\pom\Lb Q_T\Rb\,\,=\,\,1\,\,+\Delta^{(2)}\,\,-\,\,\alpha'^{(2)}_\pom \,Q^2_T
  \eeq
  with $\Delta^{(2)} = \as/\epsilon^{3})$ ($\epsilon^{3}$ = 3.8)
  and $\alpha'^{(2)}_\pom 
  =    0.34 \frac{\as}{m^2}$.
  However, at $\omega = 0$ there exists a standing branch point.
 For $Q^2_T \to \infty$ the Regge pole approaches $\omega = 0$.
 The trajectory of \eq{QCD25} means that the scattering amplitude
 has the following form
  \beq \label{QCD26}
  \widetilde{N}\Lb Y; q, b\Rb  \,\,=\,\,e^{ \Delta^{(2)}\,Y}\Lb
 \sqrt{\frac{ \pi}{ \alpha'^{(2)}_\pom\,Y}}\,e^{ - \frac{b^2}{
 4 \alpha'^{(2)}_\pom\,Y}  }\Rb
  \eeq
  The dependence on the sizes of the scattering dipoles or on
 their transverse momenta, 
  only enters  the vertices of the interactions of this Regge pole
 with the particles. Therefore, in this theory we obtain the typical
 soft Pomeron of the parton model,  with the diffusion in $b$ but
 without random walk in $\ln \Lb p_T\Rb$.

\subsection{ The hierarchy of scales: the interface of soft and hard Pomerons. }
In the previous section we saw, that the infrared divergency could
 generate the soft Pomeron: the Reggeon with the intercept larger
 than 1 and with the trajectory of the type of \eq{QCD25}. However, such
 a soft Pomeron should exist 
simultaneously with the BFKL Pomeron, and we have to find  the high
 energy asymptotic behaviour of the scattering amplitude, taking into
 account the interface of these two Pomerons. The second problem is to
 find a new dimensional scale, which in the  spontaneously broken 2 + 1
 QCD is the coupling constant $g $, which has the  dimension of mass. 

\begin{figure}[ht]
     \begin{center}
     \includegraphics[width=0.8\textwidth]{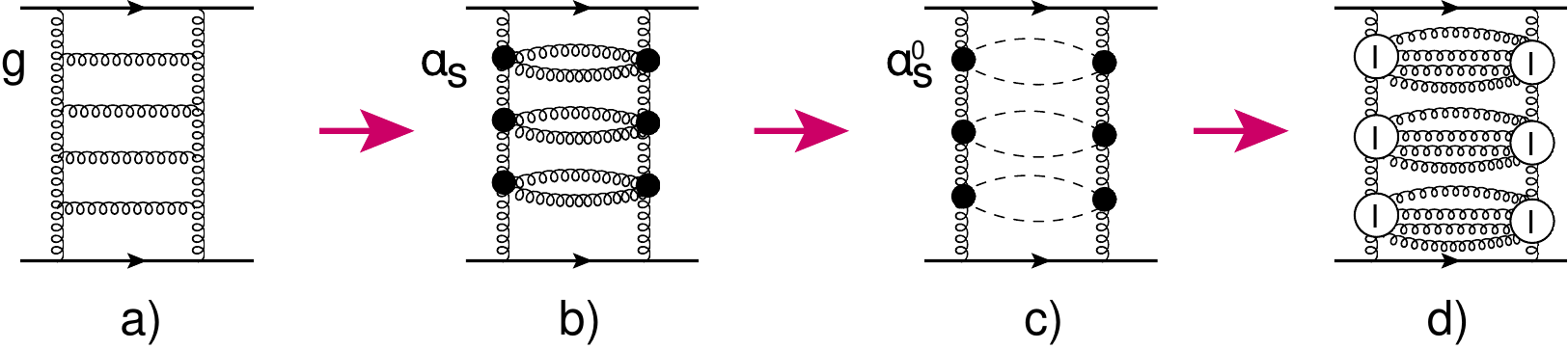} 
     \end{center}    
      \caption{Multi-peripheral (ladder) diagrams contributing to the
 leading order BFKL Pomeron(\fig{inst}-a) and to the soft Pomeron 
(\fig{inst}-b - \fig{inst}-d). The helical  lines describe the gluons,
 the dashed lines correspond to pions. \fig{inst}-d shows the
 contribution of the QCD instantons to the soft Pomeron structure. }
\label{inst}
   \end{figure}

In Ref.\cite{KHLE} it is shown  
 that such a scale exists in the QCD vacuum as a consequence 
of scale anomaly, and it is related to the density of vacuum gluon fields
 due to 
semi--classical fluctuations: numerically, $M_0^2 \simeq 4 \div 6$ GeV$^2$ .
\fig{inst}-b and \fig{inst}-c illustrate how this new scale enters the
 structure of the soft Pomeron. Indeed, among 
  the higher order, $O(\alpha^2_S)$ ($\alpha_S= g^2/4\pi$) 
corrections, to the BFKL kernel (see \fig{inst}-b).  We 
isolate a particular class of diagrams, which include the propagation 
of two gluons in the scalar color singlet channel $J^{PC}=0^{++}$.
The vertex of their production, generated 
the four-gluon coupling in the QCD Lagrangian, is $\sim \alpha_s F^{\mu\nu\ a} 
F_{\mu\nu}^a$.  We observe, that this vertex is  
proportional to the trace of the
QCD energy--momentum tensor ( $\theta^{\mu}_{\mu} $) in the chiral limit 
of massless quarks:
\begin{equation}
\theta_\mu^\mu = 
\frac{\beta(g)}{2g} F^{a\alpha\beta} F_{\alpha\beta}^{a} \simeq 
- \frac{b g^2}{32 \pi^2} F^{a\alpha\beta} F_{\alpha\beta}^{a}; \label{trace} 
\end{equation} 
The contribution of \fig{inst}-b, therefore, 
appears proportional to the correlator of the QCD energy - momentum
 tensor, for which we have 
\begin{equation}
\Pi(q^2) = i \int d^4 x\ e^{iqx}\ 
\bra{0} \T \left\{ \theta_\mu^\mu(x)\theta_\nu^\nu(0) \right\} \ket{0}
=
\int d M^2 \frac{\rho_\theta (M^2)}{M^2 - q^2-i\epsilon}
 \xrightarrow{q \to 0} -4 \langle 0|\theta^{\mu}_{\mu} |0\rangle =
 - 16 \epsilon_{\rm vac} \neq 0
\label{corr}
\end{equation}
From \eq{corr} one can see, that this contribution is not
 proportional to $\as$, and 
the scale $M_0$ is the typical mass in \eq{corr} where 
$
M_0^2 \ \ \simeq \ \ 32 \pi \left\{ {|\epsilon_{\rm vac}|
 \over N_f^2 - 1}\right\}^{\frac{1}
{2}}$.
 
Therefore,  as a consequence 
of scale anomaly,  these, apparently  $O(\alpha^2_S)$, contributions become 
the {\it dominant} ones, $O(\alpha^0_S)$.

\begin{figure}[ht]
     \begin{center}
     \includegraphics[width=0.8\textwidth]{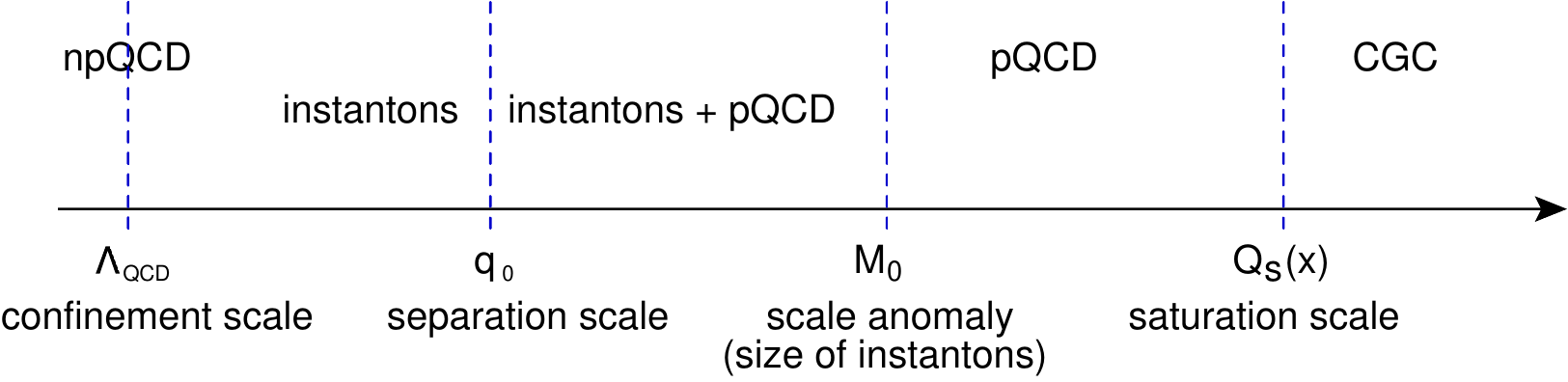} 
     \end{center}    
      \caption{Scales of QCD. }
\label{scale}
   \end{figure}

 The disappearance of the coupling constant 
in the spectral density of the $g^2 F^2$ operator can be explained, if we
 assume, that
non-perturbative QCD vacuum is dominated by the semi--classical
 fluctuations 
of the  gluon field. Since the strength of the classical gluon field 
is inversely proportional to the coupling, $F \sim 1/g$, 
the quark zero modes, and the spectral density of their pionic excitations, 
appear independent of the coupling constant. Summing the ladder
 diagrams of \fig{inst}-c we obtain the soft Pomeron with the intercept:
\beq \label{DELTA}
\Delta\,\,=\,\, \frac{1}{48}\,\,\ln \,\frac{M^2_0}{4 m^2_{\pi}}\,\,.
\eeq
In Ref.\cite{KKL} the particular type of these classical fields are considered:
we will assume, that the field is given by the instanton solution.
 It is shown, that these fields lead to the soft Pomeron (see ladder
 diagram of \fig{inst}-d). The new dimensional scale in this
 particular approach is the typical size ($\rho_0$) of the
 instanton, which comes from the lattice calculation ($\rho_0 \sim 0.3 fm$). Summing the diagrams of \fig{inst}-d we obtain the soft Pomeron contribution with the trajectory
\beq \label{HS1}
\alpha\Lb Q_T\Rb\,\,=\,\,1\,\,+\,\,\Delta_\pom\,\,-\,\,\alpha'_\pom \,Q^2_T
\eeq
with $\Delta_\pom \approx 0.08$ and $\alpha'_\pom \approx 0.25 \,GeV^{-2}$.
In \fig{scale} we plot the scales of QCD as they stem from our approach
 to the soft Pomeron of \eq{HS1}.

\begin{figure}[ht]
     \begin{center}
     \includegraphics[width=0.85\textwidth]{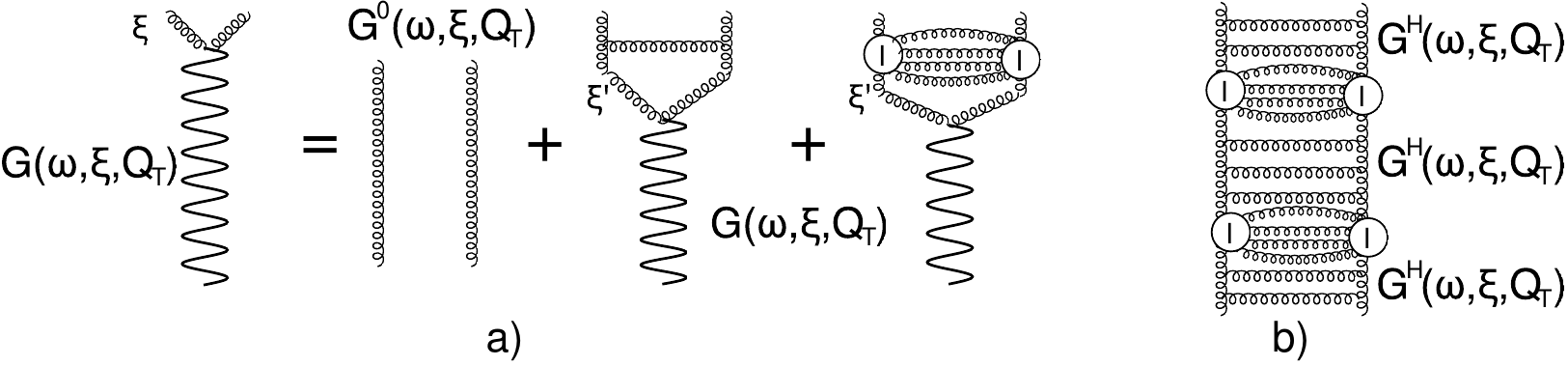} 
     \end{center}    
      \caption{Equation for the Pomeron Green function $G\Lb \omega,
 \xi,Q_T\Rb$( wavy lines in \fig{sheq}-a). $G^0\Lb \omega, \xi,Q_T\Rb$ 
denotes the Green function for the exchange of two gluons while $G^H\Lb
 \omega, \xi,Q_T\Rb$  is the Green function of the BFKL Pomeron.\fig{sheq}-b
 illustrates the solution of \eq{SOLSHEQ}. }
\label{sheq}
   \end{figure}

 Bearing in mind the hierarchy of the scales shown in  \fig{scale}, 
 one can see, that the equation for the resulting  Pomeron Green
 function takes the form\cite{BLT} (see \fig{sheq}-a)
   \bea \label{HS2}
  &&  G\Lb \omega, \vec{r}_f,\vec{r}_i,\vec{Q}_T\Rb\,\,=\\
  &&\,\, G^0\Lb \omega, \vec{r}_f,\vec{r}_i,\vec{Q}_T\Rb    \,\,+\,\,\int d^2 r'\,d^2 r"\, G^0\Lb \omega, \vec{r}_f,\vec{r}',\vec{Q}_T\Rb \Bigg\{ K^S\Lb \vec{r}', \vec{r}", Q_T\Rb \,+\,  K^{\rm BFKL}\Lb \vec{r}', \vec{r}"\Rb  \Bigg\} G\Lb \omega, \vec{r}",\vec{r}_i,\vec{Q}_T\Rb
 \eea  
    In \eq{HS2} $K^{\rm BFKL}\Lb \vec{r}', \vec{r}"\Rb$ is familiar
 kernel of the BFKL equation (see \eq{KERC}), while $K^S$  denotes
the
 kernel of the soft Pomeron which we take in the form:
    \beq \label{HS3}
    K^S\Lb \vec{r}', \vec{r}", Q_T\Rb\,\,=\,\,\Delta\Lb Q_T\Rb \,D\Lb r'\Rb\,D\Lb r"\Rb
    \eeq
    with  $ \Delta\Lb Q_T\Rb\,\,=\,\,\Delta_\pom\,\,-\,\,\alpha'_\pom
 \,Q^2_T  $(see \eq{HS1} ) and $D\Lb r\Rb$ is the function, which
 decreases with $r $ for $r > r_{\rm soft} = 1/M_0$, where $M_0\,
 \approx 2 \,GeV$, as we have discussed above. In addition we normalize
 $\int d^2 r\, D\Lb r\Rb\,=\, 1$.

      The solution to this equation, which is clear from \fig{sheq}-b, has
 the form (see appendix B of Ref.\cite{BLT}):
      \bea \label{SOLSHEQ}
 &&   G\Lb \omega, \vec{r}_f,\vec{r}_i,\vec{Q}_T\Rb\,=\\
 &&\, G^H\Lb \omega, \vec{r}_f,\vec{r}_i,\vec{Q}_T\Rb\,+\,\Delta\Lb Q_T\Rb \frac{\int d^2 r' D\Lb r'\Rb  G^H\Lb \omega, \vec{r}_f,\vec{r}',\vec{Q}_T\Rb \,  \int d^2 r" D\Lb r"\Rb  G^H\Lb \omega, \vec{r}",\vec{r}_i,\vec{Q}_T\Rb }{1\,-\,  \Delta\Lb Q_T\Rb \int \int d^2 r'\,d^2 r"\, D\Lb r'\Rb\, G^H\Lb \omega, \vec{r}',\vec{r}",\vec{Q}_T\Rb\, D\Lb r"\Rb}\nn
\eea    
        
    The new singularities of the Green function stem  from the equation:    
   \beq \label{HS4} 
    1\,-\,  \Delta\Lb Q_T\Rb \int \int d^2 r'\,d^2 r"\, D\Lb r'\Rb\, G^H\Lb \omega, \vec{r}',\vec{r}",\vec{Q}_T\Rb\, D\Lb r"\Rb   \,\,=\,\,0
    \eeq 
Integrals over $r'$ and $r"$ lead to $r' = r''\approx r_{\rm soft}$, and  
 the Green function of the BFKL Pomeron (see \eq{GF1} and \eq{GF2})  is
 equal to $\frac{1}{2 \sqrt{D \Lb\omega - \omega_0\Rb}}$ . Therefore,
 \eq{HS4} reduces to
\beq \label{HS5}
4\,D\Lb \omega - \omega_0\Rb = \Delta^2\Lb Q_T\Rb~~~~~~~~~\mbox{or}~~~
 \omega = \omega_0 + \frac{\Delta^2\Lb Q_T\Rb}{4\,D}\,= 
\,\frac{\Delta^2_\pom}{4\,D}
 \,-\,\frac{\alpha'_\pom}{2 D}\,\Delta_{\pom}\,Q^2_T
  \eeq   
    
    From \eq{HS5} we see, that the resulting intercept increases, 
 which means that the soft Pomeron influences the energy behaviour
 of the deep inelastic processes. However, for this paper the most
 significant result is the impact parameter behaviour of the Green function. 
    
     Function $G_\pom\Lb \omega, Q_T; r_1, r_2\Rb$ has been found 
in Ref.\cite{LIP}, and  it takes the following form (see Eq.35 in 
Ref.\cite{LIP})
 \beq \label{GBFKL}
 G_\pom\Lb \omega, Q_T; r_1, r_2\Rb\Big{|}_{\begin{minipage}{1.5cm}{$\omega \to \omega_0\\ 
 Q_T \to 0$ }\end{minipage}}\,\,=\,\,\,\frac{\pi}{\kappa_0}\Bigg\{ \Lb \frac{r_1}{r_2}\Rb^{\kappa_0}\,\,+\,\, \Lb \frac{r_2}{r_1}\Rb^{\kappa_0}\,\,-\,\,2\,\Lb Q^2_T\,r_1\,r_2\Rb^{\kappa_0}\Bigg\}
  \eeq     
    From \eq{GBFKL} one can see, that in the kinematic region  of
 small $Q_T$ ,where $|(\omega - \omega_0)\,\ln^2\Lb Q^2_T r_1
 r_2\Rb|\,\gg\,1$  $  G_\pom\Lb \omega, Q_T; r_1, r_2\Rb\,\,\to\,
\, 1/\sqrt{\omega - \omega_0}$ ( see section II-B-3), while in the
 region of $|(\omega - \omega_0)\,\ln^2\Lb Q^2_T r_1 r_2\Rb|\,\sim
\,1$, $G_\pom$ is less singular ($G_\pom\Lb \omega, Q_T; r_1, 
r_2\Rb\,\,\to\,
\, \sqrt{\omega - \omega_0}$.    We will restrict ourselves to the region 
of
 small $Q_T$, since we are interested in the large impact parameter
 dependence of the Green function.

    It can be estimated from the second term of \eq{SOLSHEQ},
 which takes the form
    \bea \label{HS6}
    G\Lb Y, \vec{r}_f,\vec{r}_i,\vec{b}\Rb \,&=&\,\int^{\epsilon + i \infty}_{\epsilon - i \infty} \frac{d \omega}{ 2\,\pi\,i}\int \frac{d^2 Q_T}{4\, \pi^2}e^{ \omega_0 \,Y + \delta\omega Y\,+\,\kappa_0\Lb \xi_f + \xi_i\Rb\,\,+\,\,\,i \vec{Q}_T \cdot \vec{b} } \,\frac{1}{ 2  \sqrt{D \,\delta \omega} - \Delta_\pom + \alpha'_\pom Q^2_T}\,\nn\\
    &=&\,\int^{\epsilon + i \infty}_{\epsilon - i \infty} \frac{d \omega}{ 2\,\pi\,i}   \frac{1}{2\,\pi}  e^{\omega_0 Y\,+\,\kappa_0\Lb \xi_f + \xi_i\Rb} e^{\delta \omega\,Y}\,K_0\Lb b\, \sqrt{\frac{\sqrt{\delta \omega} - \tilde{\Delta}_\pom}{4\,\alpha'_\pom \,D}}\Rb
    \eea
    where $\xi_f \,=\,\ln\Lb r^2_f/r^2_{\rm soft}\Rb$ and $\xi_i\,=\,\ln\Lb r^2_i/r^2_{\rm soft}\Rb$, $\delta \omega = \omega - \omega_0$,   $\tilde{\Delta}_\pom = \Delta_\pom/\sqrt{4 D}$ and $\kappa_0 = \sqrt{\delta \omega/D}$.
    
        Introducing a new variable $t = \sqrt{\delta \omega} -
 \Delta_\pom$ we can re-write \eq{HS6} in the form:
       \bea \label{HS7}
    G\Lb Y, \vec{r}_f,\vec{r}_i,\vec{b}\Rb \,&=&\,\int^{\epsilon + i \infty}_{\epsilon - i \infty} \frac{2(t+\Delta_\pom)\,d t  }{ 2\,\pi\,i}   e^{ \omega_0  \,Y} \exp\Lb \Lb t +\tilde{\Delta}_\pom\Rb^2 \,Y \,\,+\,\,\frac{t + \tilde{\Delta}_\pom}{\sqrt{D}}\Lb \xi_f + \xi_i\Rb\,\,- \,\, b\, \sqrt{ \frac{t}{ \sqrt{4\, \alpha'^2_\pom\, D}}}\,\,\Rb\nn\\
    &=& \,\int^{\epsilon + i \infty}_{\epsilon - i \infty} \frac{2 (t+ \tilde{\Delta}_\pom)\,d t  }{ 2\,\pi\,i}   e^{ \omega_0  \,Y} \exp\Lb \Lb t + \tilde{\Delta}_\pom\Rb^2 \,Y \,\,+\,\,\Lb t + \tilde{\Delta}_\pom\Rb\Lb \tilde{\xi}_f + \tilde{\xi}_i\Rb\,\,- \,\, b\, \sqrt{ \frac{t}{ \tilde{\alpha}'_\pom}}\,\,\Rb    \eea
    where $\tilde{\xi}_i=\xi/\sqrt{D}$ and $  \tilde{\alpha}'_\pom = 2 \sqrt{D} \alpha'_\pom$.

    Taking the integral of \eq{HS7} using the method of steepest
 descent, we obtain the  following equation for 
    the saddle point value:
    \beq \label{HSSP}
    2 \Lb t_{SP}\,+\,\tilde{\Delta}_\pom\Rb\,Y\, \,+\,\,\Lb\tilde{ \xi}_f + \tilde{ \xi}_i\Rb   \,\, - \,\,\frac{b}{2 \sqrt{t_{SP}\,\tilde{\alpha}'_\pom }}\,\,=\,\,0
    \eeq
     For large $b$ : $\frac{b}{2 \sqrt{t_{SP}\,\tilde{\alpha}'_\pom }} \,\gg\,\Lb\tilde{ \xi}_f + \tilde{ \xi}_i\Rb$, and $t_{SP} \gg \tilde{\Delta}_\pom$  the solution to \eq{HSSP} is 
    $t = t_{SP} = \Lb \frac{b}{4 \sqrt{\tilde{\alpha}'_\pom}\,\, Y}\Rb^{2/3}$.  
    
    Substituting $t_{SP}$ into  the conditions for large $b$ we obtain:
    \beq \label{LB1}
   \frac{ b}{\sqrt{\tilde{\alpha}'_\pom}} \,\gg\,\tilde{\Delta}^{3/2}_\pom 4 \,Y;~~~~~~~~~
   \frac{b^2}{\tilde{\alpha}'_\pom}\,\gg\,2 \frac{\Lb\tilde{ \xi}_f + \tilde{ \xi}_i\Rb^3}{Y}
   \eeq

    Plugging this $t_{SP}$ into \eq{HS7} one can see that 
    \bea \label{HS8}
     && G\Lb Y, \vec{r}_f,\vec{r}_i,\vec{b}\Rb   \,\,\propto \,\,\,\\
      && \exp\Lb\Lb \omega_0 + \tilde{\Delta}^2_\pom\Rb \,Y\,\,+2 \tilde{\Delta}_\pom\,Y\,\Lb \frac{b^2}{16\,\tilde{\alpha}'_\pom\,Y^2}\Rb^{1/3}\,\,+\,\,\Lb\Lb \frac{b^2}{16\,\tilde{\alpha}'_\pom\,Y^2}\Rb^{1/3}\,\,+\,\,\tilde{\Delta}_\pom\Rb \Lb \xi_f + \xi_i\Rb\,\, - \,\,\frac{3}{4} \Lb \frac{b^4}{4\tilde{\alpha}'^2_\pom \,\,Y}\Rb^{1/3}\Rb\nn
      \eea    
     .
      
   Two lessons that  we can learn from \eq{HS8}: (i)
 the random walk in $b$ leads to $\VEV{b^2}\,\,
 \propto\,\,2\tilde{\alpha}'_\pom\, \sqrt{Y}$ and
 (ii) that the eikonal  unitarization leads to the
 Froissart disc with radius 
   \beq \label{HSR}
   R\, \, \propto\,\,  \Lb \omega_0 + \tilde{\Delta}_\pom^2 + 2 \Lb \frac{4}{3}\Lb \omega_0 + \tilde{\Delta}^2_\pom\Rb^{3/2}\Rb\Rb \,Y\,\,+\,\, \Lb\Lb \frac{4}{3}\Lb \omega_0 + \tilde{\Delta}^2_\pom\Rb\Rb^{3/2} \,+\,\tilde{\Delta}_\pom\Rb\,\Lb \xi_f + \xi_i\Rb.\eeq 
   One can see that \eq{HSR} satisfy the conditions of \eq{LB1}.
  Note, that we obtain the value of $\VEV{b^2}$ 
 and the radius of the Froissart disc in accord with the
 general analysis in section II-B-4.    
\subsection{Summing pion loops}
\begin{figure}[ht]
     \begin{center}
     \begin{tabular}{c c  c}
     \includegraphics[width= 0.1\textwidth]{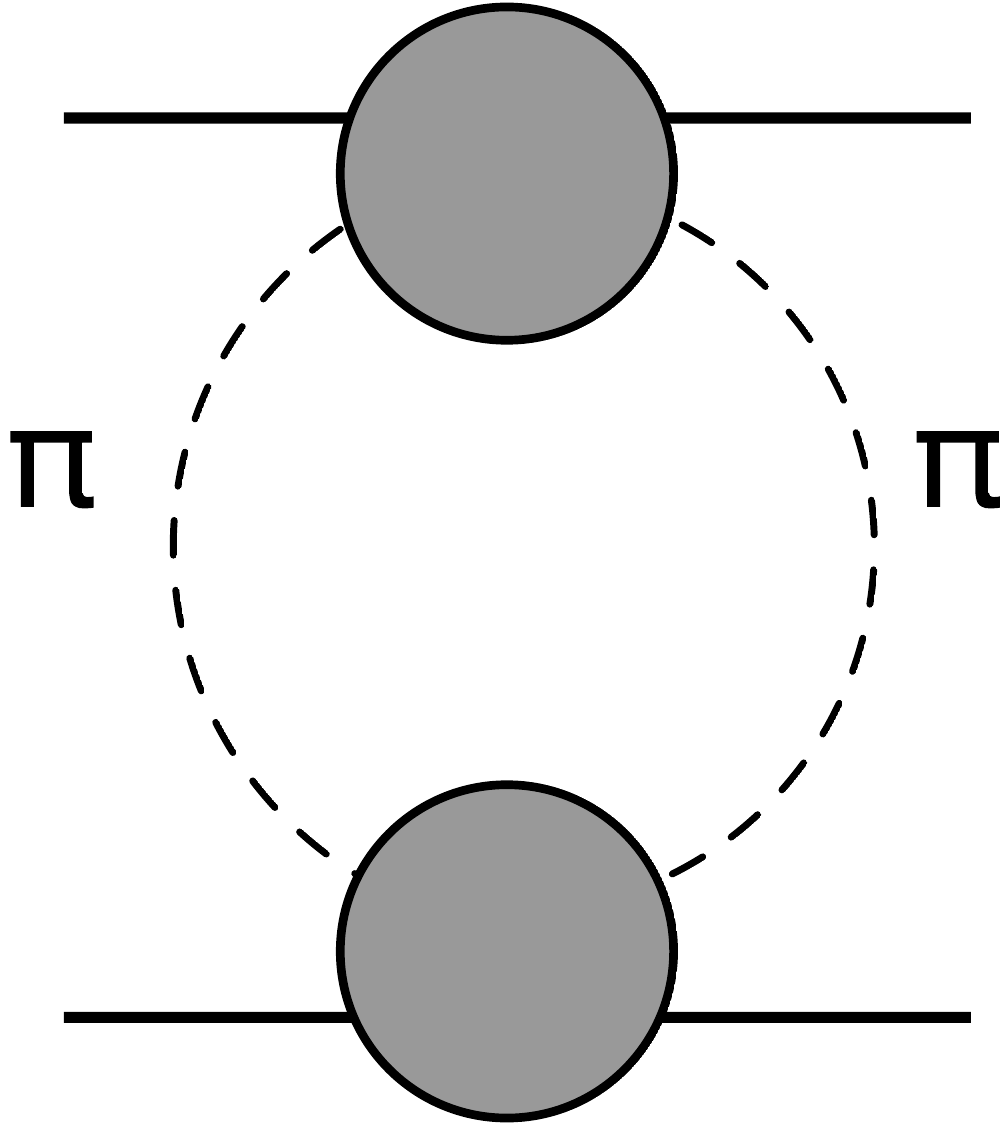} &   ~~~~~~~~~~~~~~~~~~~~~&\includegraphics[width=0.25\textwidth]{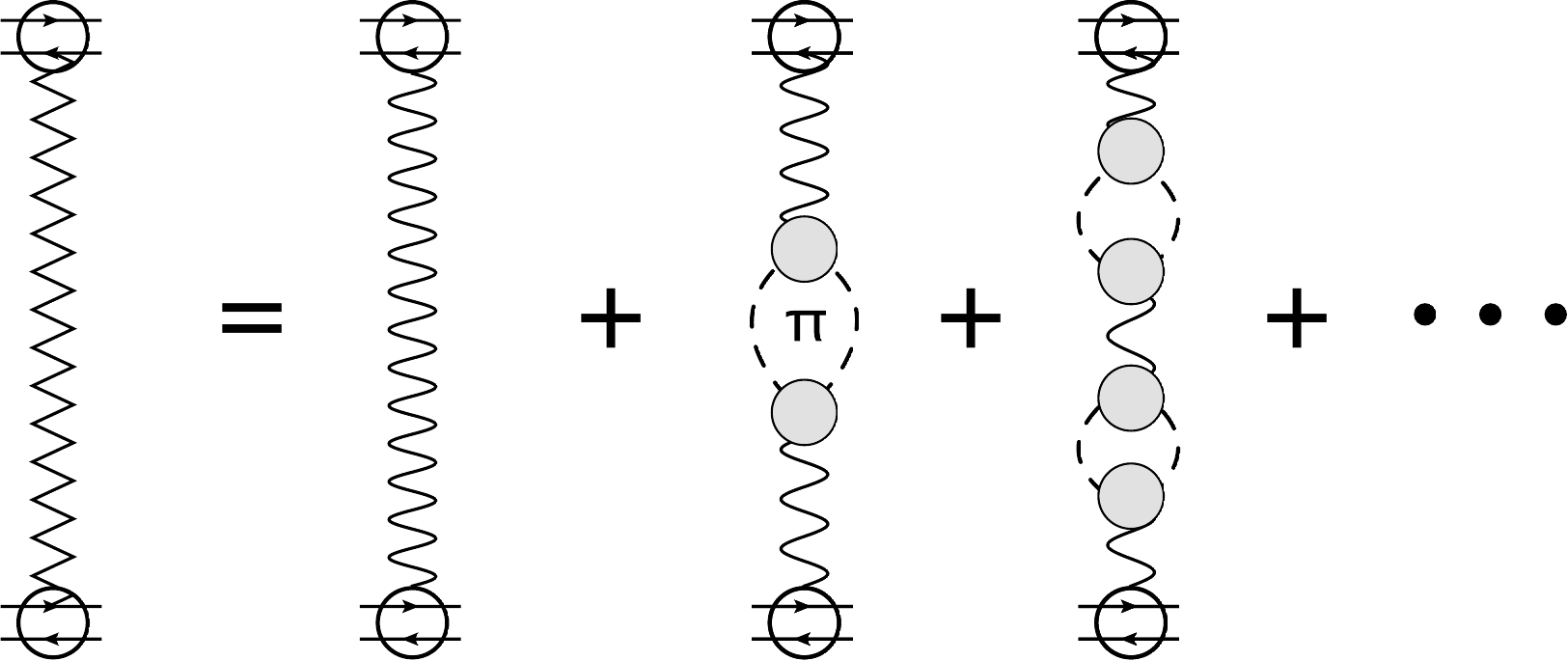} \\
     \fig{2pi}-a & &  \fig{2pi}-b\\
     \end{tabular}
       \end{center}    
      \caption{\fig{2pi}-a: The exchange of two pions, that gives the
 main contribution to the large $b$ behaviour of the scattering
 amplitude\protect\cite{FROI}.\fig{2pi}-b:The diagrams that
 describe the pion loop contribution to the BFKL Pomeron. The
 wavy line describe the BFKL Pomeron, the dashed line stands
 for pion. The zigzag line denotes the resulting Pomeron.}
\label{2pi}
   \end{figure}

        In Ref.\cite{LEPION}  a theoretical approach  is
 proposed, which is based  on the assumption, that the BFKL
 Pomeron, being perturbative in nature, takes into account
 rather short distances (say of the order of $1/m_G$, where
 $m_G$ is the mass of the lightest glueball); and the long 
distance contribution can be described by the exchange of
 pions ($1/2\mu\,\gg\,1/m_G$, where $\mu$ is the pion mass).    Indeed, 
 the proof of the Froissart theorem indicates, that the
 exponential fall-off at large $b$ is closely related
 to the contribution of the exchange of two lightest hadrons: 
 2 pions, in $t$-channel (see \fig{2pi}-a).  \fig{2pi}-b demonstrates
 both the main idea and the 
 practical method of our approach. The set of diagrams of
 \fig{2pi} can be summed, using the following equation:
 \beq \label{PIONEQ}
G_\pom\Lb \omega, Q_T; R_\pi,R_\pi\Rb \,\,=\,\,G^{\rm BFKL}_\pom\Lb \omega, Q_T; R_\pi,R_\pi\Rb\,\,\,+\,\,G^{\rm BFKL}_\pom\Lb \omega, Q_T; R_\pi,R_\pi\Rb \,\Sigma\Lb \omega, Q_T\Rb
G_\pom\Lb \omega, Q_T; R_\pi,R_\pi\Rb
\eeq 
where $G_\pom$ denotes the resulting Green's function
 of the Pomeron with the pion loops for $r_1=r_2 = R_\pi$
 where $R_\pi$ is the size of the pion. $G^{\rm BFKL}_\pom$
 is the Green function of the BFKL Pomeron.

\beq \label{PIONSOL}
G_\pom\Lb \omega, Q_T; R_\pi,R_\pi\Rb \,\,=\,\,\frac{1}{G^{-1}_\pom\Lb \omega, Q_T; R_\pi,R_\pi\Rb\,\,-\,\,
\Sigma\Lb \omega, Q_T\Rb}
\eeq
The $t$-channel unitarity for $G_\pom\Lb \omega, Q_T; R_\pi,R_\pi\Rb$ is
 used to calculate $\Sigma$,  as it was suggested in Ref.\cite{ANGR}. 
 The unitarity constraints for $G_\pom$ takes the form \cite{COL,GRLEC}
 ($t = - Q^2_T$):
\beq \label{PIONUNIT}
-i \{ G_\pom\Lb \omega, t + i\epsilon ; R_\pi, R_\pi\Rb \,\,-\,\,G_\pom\Lb \omega, t - i\epsilon ; R_\pi, R_\pi\Rb  \}\,\,=\,\,\rho\Lb \omega, t \Rb G_\pom\Lb \omega, t + i\epsilon ; R_\pi, R_\pi\Rb\,G_\pom\Lb \omega, t - i\epsilon ; R_\pi, R_\pi\Rb
\eeq
\eq{PIONUNIT} is written for $ t \,>\, 4 \mu^2$ where $\mu$ is the mass
 of pion. We wish to only sum  the  two pion contribution,
assuming that all other  have been taken into account in the BFKL Pomeron.
 In other words, we believe that \eq{PIONUNIT} can be trusted for $t
 \,<\,m^2_G$ where $m_G$ is the lightest glueball. $\rho \Lb \omega, t\Rb$
 is equal to $( t - 4 \mu^2)^{\frac{3}{2} + \omega}/\sqrt{t}$.
Plugging in \eq{PIONUNIT} the solution of \eq{PIONSOL}  we
 obtain the following equation for $\Sigma$:
\beq \label{PIONEQSIG}
\mbox{Im}_t \Sigma\Lb \omega,t\Rb\,\,=\,\,\h \rho \Lb \omega, t\Rb\,\Big(\Gamma^\pom_{\pi \pi}\Big)^2\,\frac{1}{\omega + 2}
\eeq
where vertex $\Gamma^\pom_{\pi \pi}$ describes the interaction of the
 BFKL Pomeron with $\pi$-meson and it has been evaluated in Ref.\cite{LEPION}:
$\Gamma^\pom_{\pi \pi}\,\,=\,\,=\,\,\frac{8}{9}\,\bas\,R_\pi$.
Experimentally $< R^2_\pi> = 0.438 \pm 0.008 \,fm^2$ \cite{PIR} and
we will bear this value  in mind
   for the numerical estimates.

We use  dispersion relations to calculate $\Sigma$. Since we believe,
 that non-perturbative corrections will not change the intercept of the
 BFKL Pomeron (see Refs.\cite{LETA,LLS}) we write the dispersion relation
 with one subtraction. It takes the form
\beq \label{DISRE1}
\Sigma\Lb \omega,t\Rb\,\,=\,\, t\, \frac{1}{\pi} \int \frac{ d t' \,\mbox{Im}_t \Sigma\Lb \omega,t'\Rb}{ t' \,(t' - t)}\,\,=\,\,t\frac{3}{2}\,\frac{1}{\omega + 2}\frac{\Lb \Gamma^\pom_{\pi \pi} \Rb^2}{  \pi} \int_{4 \mu^2}^{m_G^2} \frac{d t'  \,\rho \Lb \omega, t'\Rb}{ t' \,(t' - t)}\,\,=\,\,\bas^2\,\frac{16}{9}\,R^2_\pi\,\,\frac{1}{\omega + 2}\,\,t\int_{4 \mu^2}^{m_G^2} \,\frac{d t'  \,\rho \Lb \omega, t'\Rb}{ t' \,(t' - t)}\eeq
Factor $3/2$ stems from two states in $t$-channel: $\pi^+\,\pi^-$
 and $\pi^0\,\pi^0$.
         The integral over $t'$ for $\omega\, \ll\,1$ has been taken
 in Ref.\cite{ANGR}. For fixed $\omega$
\beq \label{DISRE2}
\Sigma\Lb \omega,Q_T\Rb\,\,\,=\,\,\,\bas^2\,\frac{16}{9}\,R^2_\pi\,\,\frac{1}{\omega + 2}\,\,(- Q^2_T)\,\frac{\Lb m^2_G\Rb^{5/2 + \omega}}{ 8 \mu^2 *\Lb 5 + 2 \omega\Rb \Lb 4 \mu^2 + Q^2_T\Rb}\,F_1\Lb \frac{5}{2} + \omega, \frac{3}{2}, 1, \frac{7}{2} + \omega, -\frac{m^2_G}{4 \mu^2}, -\frac{m^2_G}{4 \mu^2 + Q^2_T}\Rb
\eeq
where $F_1$ is the Appell $F_1$ function ( see Ref.\cite{RY} formulae
 {\bf 9.180 - 9.184}).

The linear term in $Q^2_T$ can be easily found since two arguments in
 $F_1$ coincide, and we can use the relation (see Ref.\cite{RY} formula
 {\bf 9.182(11)})  and reduce \eq{DISRE2} to the form
\beq \label{DISRE3}
\Sigma\Lb \omega,Q_T\Rb\,\,\,=\,\,\,\bas^2\,\frac{16}{9}\,R^2_\pi\,\,\frac{1}{\omega + 2}\,\,(- Q^2_T)\,\frac{\Lb m^2_G\Rb^{5/2 + \omega}}{ 8 \mu^2 *\Lb 5 + 2 \omega\Rb \Lb 4 \mu^2 + Q^2_T\Rb}\,{}_2F_1\Lb \frac{5}{2} + \omega, \frac{5}{2},  \frac{7}{2} + \omega, -\frac{m^2_G}{4 \mu^2}\Rb
\eeq
For $m_G \,\gg\,\mu$ \eq{DISRE3} takes the form
\bea \label{DISRE4}
\Sigma\Lb \omega,Q_T\Rb&=&\,\,\,\bas^2\,\frac{16}{9}\,R^2_\pi\,\,\frac{1}{\omega + 2}\,\,(- Q^2_T)\\
&\times&\Bigg\{\frac{1}{\sqrt{\pi}\, 3 \Lb 5 + 2 \omega\Rb}
4^{1+\omega} \Lb \frac{1}{\mu^2}\Rb^{\h + \omega}\!\!\!\!\!\!\mu\, \Lb 2 + 3 \omega + \omega^3\Rb \Gamma\Lb - 2 - \omega\Rb \Gamma\Lb \frac{7}{2} + \omega\Rb\,
+\,\frac{\Lb m^2_G\Rb^\omega}{\omega}\,\frac{4 + 2 \omega + 3 \omega^2 + \omega^3}{4 ( 2 + \omega)}\Bigg\}\nn\\
&=& -Q^2_T \, \alpha'_{eff}\Lb \bas\Rb\nn
\eea

We evaluate  $\alpha'_{eff}\Lb \bas\Rb$ using $R^2_\pi = 0.436  fm^2$ and 
the value for $m^2_G = 5 \,GeV^2$ that stems from lattice calculation of 
the glueball masses \cite{GLUETR} for the Pomeron trajectory.
  In principle we can consider this coefficient as the only
 non-perturbative input that we need, but it is pleasant to realize that
 simple estimates give a sizable value of $0.1\,GeV^{-2}$ for $\bas
\ = 0.3$ (see Ref.\cite{LEPION}).
  
  Comparing \eq{PIONSOL} with \eq{HS6} we see that the difference is  only
 that in \eq{PIONSOL} we put $\Delta_\pom=0$, and the new dimensional 
scale is determined by the mass  of the lightest glueball $m_G$. Bearing
 this in mind and using \eq{HS8} we obtain:
    \beq \label{PIONG}
  G\Lb Y, \vec{r}_f,\vec{r}_i,\vec{b}\Rb   \,\,\propto \,\,\,
     \exp\Lb \omega_0 \,Y\,+\,\,\Lb \frac{b^2}{16\,\tilde{\alpha}'_\pom\,Y^2}\Rb^{1/3} \Lb \tilde{\xi}_f + \tilde{\xi}_i\Rb\,\, - \,\,\frac{3}{4} \Lb \frac{b^4}{4\tilde{\alpha}'^2_\pom \,\,Y}\Rb^{1/3}\Rb
      \eeq    
      with $r_{\rm soft} = 1/m_G$.
      
  In \eq{PIONG} we  replace  \eq{PIONEQ} by  a more general \eq{HS2} with
 the solution of \eq{SOLSHEQ}.    We believe that \eq{SOLSHEQ} gives
 the solution to the problem of large $b$ behaviour.

  As we have mentioned, obtaining this equation we used, based on the
 results of Refs.\cite{LETA,LLS},  that
 the non-perturbative corrections do not change the intercept of the
 BFKL Pomeron. One  way to check this key assumption is  to
 try to describe the experimental data for the soft interaction at
 high energy, assuming, that we have only the BFKL Pomeron. In
 Refs.\cite{GLP} the model for the soft interaction was built,
 which based on two main ingredients: (i) CGC approach  for the
 scattering amplitude of two dipoles; and (ii) phenomenological
 description of the hadron structure. 
 In this model, we successfully describe the DIS data from HERA,
 the total, inelastic, elastic and diffractive cross sections,
 the t-dependence of these cross sections, as well as the
 inclusive production and rapidity and angular correlations
 in a wide range of energies, including that of the LHC.

\subsection{Resume}
       Concluding this brief review,  we believe, that \eq{PIONG}
 incorporates correctly the QCD random walk in the impact parameters
 ($b$) into the BFKL equation. We would like to emphasize, that two
 main ingredients  formed the basis of this conclusion: the assumption,
 that the non-perturbative modification of the BFKL Pomeron does not
 change its intercept; and  our result of section II-B-4, that 
 $\VEV{b^2}$ for the QCD diffusion in $b$  is proportional to $\sqrt{Y}$. 
       
Insufficiency of \eq{PIONG} is,  that it gives the modification
 of the Green function of the perturbative BFKL Pomeron,  but it
 is not written as an  alternative for the non-linear 
Balitsky-Kovchegov
 equation.    On the other hand, for discussion of the asymptotic
 behaviour of the scattering amplitude at high energy, we need a
 new evolution equation, which will include on the same footing
 both the shadowing corrections, and the correct behaviour at
 large impact parameter.     

\section{A new non-linear evolution equation}
  We proposed the following generalization of the Balitsky-Kovchegov
 non-linear evolution equation:
\bea \label{NEEQ}
\frac{\partial}{\partial Y} N\Lb \vec{r}, \vec{b} ,  Y \Rb &=&\bas\!\! \int \frac{d^2 \vec{r}'}{2\,\pi}\,K\Lb \vec{r}', \vec{r} - \vec{r}'; \vec{r}\Rb \Bigg\{N\Lb \vec{r}',\vec{b} - \h \Lb \vec{r} - \vec{r}' \Rb, Y\Rb + 
N\Lb\vec{r} - \vec{r}', \vec{b} - \h \vec{r}', Y\Rb \,\,- \,\,N\Lb \vec{r},\vec{b},Y \Rb\nn\\
& & ~~~~~~~- N\Lb\vec{r} - \vec{r}', \vec{b} - \h \vec{r}', Y\Rb   \, N\Lb \vec{r}',\vec{b} - \h \Lb \vec{r} - \vec{r}' \Rb, Y\Rb\Bigg\}\,\,\,+\,\,\,\Lb \alpha'_{\rm eff}\,\nabla^2_b\Rb^2 \,N\Lb \vec{r},\vec{b},Y \Rb\nn\\
&\xrightarrow{b \gg r,r'} & \bas\!\! \int \frac{d^2 \vec{r}'}{2\,\pi}\,K\Lb \vec{r}', \vec{r} - \vec{r}'; \vec{r}\Rb \Bigg\{N\Lb \vec{r}',\vec{b} , Y\Rb + 
N\Lb\vec{r} - \vec{r}', \vec{b} , Y\Rb \,\,- \,\,N\Lb \vec{r},\vec{b},Y \Rb\nn\\
& &~~~~~~~~~~~ - N\Lb\vec{r} - \vec{r}', \vec{b}, Y\Rb   \, N\Lb \vec{r}',\vec{b} , Y\Rb\Bigg\}\,\,\,+\,\,\,\Lb \alpha'_{\rm eff}\,\nabla^2_b\Rb^2 \,N\Lb \vec{r},\vec{b},Y \Rb\nn
\eea
  Where $\alpha'_{\rm eff}$ is a new dimensional scale. We believe that
 \eq{NEEQ}  is  the correct  way to introduce this scale in the 
non-linear
 evolution.
       \begin{boldmath}
\subsection{The BFKL Pomeron with diffusion in $b$}
\end{boldmath}

 First, let us solve the new equation at large $b$, where only the  linear
 part contributes.
 The equation takes the form:
 \beq \label{NEEQ1}
\frac{\partial}{\partial Y} N\Lb \vec{r}, \vec{b} ,  Y \Rb = \bas\!\! \int \frac{d^2 \vec{r}'}{2\,\pi}\,K\Lb \vec{r}', \vec{r} - \vec{r}'; \vec{r}\Rb \Bigg\{N\Lb \vec{r}',\vec{b} , Y\Rb + 
N\Lb\vec{r} - \vec{r}', \vec{b} , Y\Rb  -  N\Lb \vec{r}, \vec{b} ,  Y \Rb \Bigg\}\,\,\,+\,\,\,\Lb \alpha'_{\rm eff}\,\nabla^2_b\Rb^2 \,N\Lb \vec{r},\vec{b},Y \Rb
\eeq 

We can obtain the solution to this equation using  it in the
 form: $N\Lb \vec{r}, \vec{b} ,  Y \Rb\,\,=\,\,G\Lb \vec{r}, Y
 \Rb\,\Phi\Lb Y; b\Rb$ where $G\Lb \vec{r}, Y \Rb$ is the Green's
 function of the BFKL equation and $\Phi\Lb Y; b\Rb$ satisfies
 the following equation
 \beq \label{NEEQ11}
\frac{\partial}{\partial Y} \Phi\Lb \vec{b} ,  Y \Rb\,\,=\,\,\,\,\,\Lb \alpha'_{\rm eff}\,\nabla^2_b\Rb^2 \,\Phi\Lb \vec{b},Y \Rb
\eeq
In $\omega$-representation:
\beq \label{NEEQ12}
 \Phi\Lb \vec{b} ,  Y \Rb\,\,=\,\,\int^{\epsilon + i \infty}_{\epsilon - i \infty} \frac{d \,\omega}{ 2\, \pi \,i}\,e^{\omega\, Y} \, \Phi\Lb \vec{b} ,  \omega \Rb  
\eeq
\eq{NEEQ11} takes the form
\beq \label{NEEQ13}
\Lb \alpha'_{\rm eff}\,\nabla^2_b\Rb^2 \,\Phi\Lb \vec{b},\omega \Rb\,\,=\,\,\omega 
 \,\Phi\Lb \vec{b},\omega \Rb
 \eeq
 The general solution to \eq{NEEQ13} can be written
 (see Ref. \cite{MATH} formula {\bf 9.4.3}) as
 \beq \label{NEEQ14}
 \Phi\Lb \vec{b},\omega \Rb\,\,=\,\,\Phi_-\Lb \vec{b},\omega \Rb \,\,+\,\,\Phi_+\Lb \vec{b},\omega \Rb 
 \eeq
 where $\Phi_{\pm}$ are the solution to the following equations:
  \beq \label{NEEQ15} 
 \alpha'_{\rm eff}\,\nabla^2_b \,\Phi_{\pm}\Lb \vec{b},\omega \Rb \,\,
 \pm\,\,\sqrt{\omega}\,\Phi_{\pm}\Lb \vec{b},\omega \Rb\,\,=\,\,0
 \eeq
 
 We restrict ourself  by the solution $\Phi\Lb \vec{b},\omega \Rb\,\,=\,\,\,
 \Phi_+\Lb \vec{b},\omega \Rb $ for which the Green's function is equal to
   \beq \label{NEEQ15}  
 \Phi_+\Lb \vec{b},\omega \Rb\,\,=\,\,\int \frac{d^2 Q_T}{ (2 \pi )^2}\,e^{ i \vec{Q}_T\cdot \vec{b}}\frac{\phi^{in}_+\Lb \omega\Rb}{ \alpha'_{\rm eff}\,Q^2_T \,\,+\,\,\sqrt{\omega} }
 \eeq
 where $\phi^{in}_+\Lb \omega\Rb$ should be found from the initial
 condition at  $Y=0$. 
 Searching for the Green's function of \eq{NEEQ15} we impose the
 initial condition: 
  \beq \label{NEEQIC}
  \Phi_+\Lb \vec{b},Y = 0  \Rb\,\,=\,\,\delta^{(2)}\Lb \vec{b}\Rb
  \eeq
  which results in $\phi^{in}_+\Lb \omega\Rb\,\,=\,\,1/\Lb 2 \sqrt{\omega}\Rb$.

   The similarities between this equation and \eq{HS6} and \eq{PIONSOL}
 are obvious.
 Taking integral over $d^2 Q_T$ we obtain:
 \beq \label{NEEQ16}
  \Phi_+\Lb \vec{b},\omega \Rb\,\,= \frac{1}{ 4\, \pi\, \alpha'_{\rm eff}\,\sqrt{\omega}} K_0\Lb \frac{b}{\sqrt{ \alpha'_{\rm eff}}}\,\omega^{1/4}\Rb
  \eeq
 and for $ \Phi\Lb \vec{b}, Y \Rb $ we have
 \beq \label{NEEQ17}
  \Phi\Lb \vec{b}, Y \Rb\,\,=\,\,\,\int^{\epsilon + i \infty}_{\epsilon - i \infty} \frac{d \,\omega}{8 \pi^2\,i}\,\frac{1}{\, \alpha'_{\rm eff}\,\sqrt{\omega}}\,e^{\omega \,Y} \, K_0\Lb \frac{b}{\sqrt{ \alpha'_{\rm eff}}}\, \omega^{1/4}\Rb
  \eeq
   For large $b$ we can take the integral over $\omega$  plugging the
 asymptotic behaviour of the $K_0$ function and using the method of
 steepest descend. \eq{NEEQ17} takes the form:
   \beq \label{NEEQ18}
  \Phi\Lb \vec{b}, Y \Rb\,\,=\,\,\,\int^{\epsilon + i \infty}_{\epsilon - i \infty} \frac{d \,\omega}{8 \pi^2\,i}\,\frac{1}{\alpha'_{\rm eff}\,\sqrt{\omega}}\,e^{\omega \,Y} \,\sqrt{\frac{\pi}{2}} \Lb \frac{1}{\omega}\Rb^{1/8} \frac{1}{\sqrt{\frac{b}{\sqrt{ \alpha'_{\rm eff}}}}} \,e^{ - \frac{b}{\sqrt{ \alpha'_{\rm eff}}}\,\omega^{1/4}}
  \eeq  
  
  The equation for the saddle point has the following form
     \beq \label{NEEQ19}  
     Y\,\,- \,\,\frac{1}{4}\frac{\frac{b}{\sqrt{ \alpha'_{\rm eff}}}}{\omega^{3/4}_{\rm SP}}\,\,=\,\,0;~~~~~~~~~~~\omega_{\rm SP} \,\,=\,\,\Bigg( \frac{b}{ 4\,Y\, \sqrt{ \alpha'_{\rm eff}}}\Bigg)^{4/3};
        \eeq
     
    From \eq{NEEQ19} we obtain for the integral over $\omega$:
      \beq \label{NEEQ20}
  \Phi\Lb \vec{b}, Y \Rb\,\,=  \,\,\frac{ 4 \,\pi }{\sqrt{6} \, \alpha'_{\rm eff}}\Bigg(\frac{ \alpha'_{\rm eff}}{4\, Y\,b^2}\Bigg)^{1/3}\,\exp\Lb - \frac{3}{4}\,{\cal Z}\Rb
  \eeq
  where   
  \beq \label{Z}
     {\cal Z} \,=\,\Lb \frac{b^4}{4 {\alpha'}^2_{\rm eff}\,Y}\Rb^{1/3}
     \eeq   
  
Finally, from $N\Lb \vec{r}, \vec{b} ,  Y \Rb\,\,=\,\,G\Lb \vec{r}, Y
 \Rb\,\Phi\Lb Y; b\Rb$ we obtain
\beq \label{NEEQ9} 
G\Lb \vec{r}, \vec{b} ,  Y \Rb\,\,=\,\,\int^{\epsilon + i\,\infty}_{\epsilon - i\,\infty}\frac{d \gamma}{2\,\pi\,i} e^{ \bas \chi\Lb \gamma\Rb \,Y\,\,-\,\,\Lb - \gamma\Rb \xi}\,\Phi\Lb Y; b\Rb\,\,=\,\,
\,\,\frac{r}{R} e^{\omega_0\,Y}\frac{1}{2 \sqrt{\pi\,D\,Y}}\,\,e^{ - \frac{\xi^2}{4\,D\,Y}}\,\Phi\Lb Y; b\Rb
\eeq
and $\Phi\Lb Y; b\Rb$ is given by \eq{NEEQ20}. In \eq{NEEQ9} we
 integrated over
$\gamma$ using the diffusion expansion for the kernel, $\chi(\gamma)$
 for $\gamma \to \h$.

For the region of not large $b$ we obtain 
the  solution  which has a cumbersome form:
\beq \label{NEEQALLB}
\frac{1}{16\,{\alpha'}^2_{\rm eff} \,Y} \Bigg( -
 4 \sqrt{\pi}\sqrt{- {\alpha'}^2_{\rm eff} \,Y}\,\,
 _0F_2\left(;\frac{1}{2},1; -\,\frac{b^4}{256 \, {\alpha'}^2_{\rm eff}\,Y }\right)-b^2 \, _0F_2\left(;\frac{3}{2},\frac{3}{2};-\,\frac{b^4}{256 \, {\alpha'}^2_{\rm eff} \,Y}\right)\Bigg)
\eeq
where ${}_0F_2$ is the hyperbolic function (see formula 
   {\bf 9.14} of Ref.\cite{RY}).

As we have mentioned above the solution of \eq{NEEQ20} is the Green's function
 of
\eq{NEEQ11},  and it    preserves the remarkable property of the
 Green's function of the BFKL Pomeron:
\bea \label{GFUNIT0}
\Phi\Lb  \vec{b}  , Y \Rb\,\,&=&\,\,\int\,d^2 b' \,\Phi\Lb\vec{b} -\vec{b}' , Y - Y' \Rb\,\,\Phi\Lb  \vec{b}' ,  Y' \Rb\,\,\\\
&=&\,\,\,\int\,d^2 b'  \int \frac{d^2 Q'_T}{ 4 \pi^2}\exp\Bigg( \,\,{\alpha'}^2_{\rm eff}\,Q'^4_T\,Y'\,\,
+ i \vec{Q'}_T \cdot  \vec{b}' \Bigg)\, \int \frac{d^2 Q_T}{ 4 \pi^2}\exp\Bigg(\,\,{\alpha'}^2_{\rm eff}\,Q^4_T\,\Lb Y\,-\,Y'\Rb
+ i\, \vec{Q}_T \cdot \Lb \vec{b} - \vec{b}'\Rb\Bigg)\nn\\
&=&\,\,\int d^2 Q_T 
\exp\Big( {\alpha'}^2_{\rm eff}\,Q'^4_T\,\,+\,\,i \,\vec{Q'}_T \cdot  \vec{b}\Big)\eea
In \eq{GFUNIT0} we use \eq{NEEQ15}  for $\Phi_+\Lb \vec{b},\omega\Rb$
 which leads to
the following representation for $\Phi\Lb  \vec{b}  , Y \Rb$

\beq \label{GFUNIT1}
\Phi\Lb  \vec{b}  , Y \Rb\,\,=\,\,\int^{\epsilon + i\,\infty}_{\epsilon - i\,\infty} \frac{ d \omega}{2\, \pi\,i}\,e^{ \omega\,Y}\, \Phi_+\Lb \vec{b},\omega\Rb\,\,=\,\,\int d^2 Q_T 
\exp\Big( {\alpha'}^2_{\rm eff}\,Q'^4_T\,\,+\,\,i\, \vec{Q'}_T \cdot  \vec{b}\Big)
\eeq 
One can see, that the integration over $d^2 b'$ in \eq{GFUNIT0}
 generates $\delta{(2)}
\Lb \vec{Q}_T - \vec{Q}'_T\Rb$, which results in \eq{GFUNIT0}.

Using \eq{GFUNIT0} we can prove that for the Green's function of
 \eq{NEEQ9} we have the following equation:

\bea \label{GFUNIT}
G\Lb \vec{r}, \vec{b} , \vec{R} , Y \Rb\,\,&=&\,\,\int d^2 r' \,d^2 b' \,G\Lb \vec{r}, \vec{b} -\vec{b}', \vec{r}' , Y - Y' \Rb\,\,G\Lb \vec{r}', \vec{b}' , \vec{R}, Y' \Rb\,\,\\\
&=&\,\,\,\int d \xi' \,d^2 b' \int \frac{ d \gamma'}{2\,\pi\,i} \int \frac{d^2 Q'_T}{ 4 \pi^2}\exp\Bigg( \Lb \bas 
\chi\Lb \gamma'\Rb\,\,+\,\,{\alpha'}^2_{\rm eff}\,Q'^4_T\Rb\,Y'\,+\,\Lb \gamma' - 1\Rb \, \xi'
+ i \vec{Q'}_T \cdot  \vec{b}' \Bigg)\nn\\
&\times&\,\int \frac{ d \gamma}{2\,\pi\,i} \int \frac{d^2 Q_T}{ 4 \pi^2}\exp\Bigg( \Lb \bas 
\chi\Lb \gamma\Rb\,\,+\,\,{\alpha'}^2_{\rm eff}\,Q^4_T\Rb\,\Lb Y\,-\,Y'\Rb\,+\,\Lb \gamma - 1\Rb \Lb \xi - \xi'\Rb
+ i \vec{Q}_T \cdot \Lb \vec{b} - \vec{b}'\Rb\Bigg)\nn
\eea
where $\xi'\,=\,\ln\Lb r'^2/R^2\Rb$. 

The  integration over $\xi'$ and $\vec{b}'$ gives 
 $\delta\Lb \gamma  - \gamma'\Rb$ and $ \delta^{(2)}\Lb
 \vec{Q}_T \,-\,\vec{Q}'_T\Rb$ , which leads to  tequation \eq{GFUNIT}.

One can see that \eq{NEEQ9} has the same main features as \eq{PIONG}.
 In particular, the unitaristion leads to the Froissart disc with radius
\beq \label{FD1}
\frac{R}{\sqrt{{\alpha'}_{\rm eff}}}\,\,=\,\, \frac{4}{3^{3/4}}\, \omega_0^{3/4} \,Y\,\,\,-\,\,
\frac{3^{1/4}}{4\,D\, \omega^{1/4}_0} \frac{\xi^2}{Y}
\eeq
 whose edge decreases with $b$ as $e^{ - \frac{3}{4} \Lb \frac{b^4}{4 
{\alpha'}^2_{\rm eff}\,Y}\Rb^{1/3}}$ which corresponds to the diffusion
 in $b$ with $\VEV{b^2} \propto \sqrt{Y}$, as  follows from section II-B-4.
     
\subsection{Solution in the vicinity of the saturation scale}

The behaviour of the scattering dipole amplitude in the
 vicinity of the saturation scale can be found from the
 solution of the linear equation\cite{GLR,MUT}. We will
 use the solution to \eq{NEEQ1} in the form of \eq{NEEQ20},
 and will take the integral over $\gamma$ using the method
 of steepest descend. The equation for the saddle point in
 $\gamma$ has the form:
\beq \label{SOLVS1}
\bas \frac{d \chi\Lb \gamma\Rb}{ d \gamma}\Bigg{|}_{\gamma = \gamma_{\rm SP}}\,Y \,\,+\,\,\xi\,\,=\,\,0
\eeq

The second condition is that the scattering amplitude is a constant
 for $r^2\,Q^2_s = 1$, which  has the form 
\beq  \label{SOLVS2}
\bas \, \chi\Lb \gamma_{\rm SP}\Rb\,Y \,\,+\,\,\Lb \gamma_{\rm SP}\,-\,1\Rb\xi\,\,-\,\,\frac{3}{4} {\cal Z}\,\,=\,\,0
\eeq

Plugging \eq{SOLVS1} into \eq{SOLVS2} we obtain the following equation
 for $\gamma_{\rm SP}$.
\beq  \label{SOLVS3}
\bas \Lb \chi\Lb \gamma_{\rm SP}\Rb \,\,+\,\,\Lb 1 -  
\gamma_{\rm SP}\Rb \,\frac{d \chi\Lb \gamma\Rb}{ d
 \gamma}\Big{|}_{\gamma = \gamma_{\rm SP}}\Rb \,Y\,\,=\,\,\frac{3}{4} {\cal Z}
\eeq

For ${\cal Z} = 0$ the solution to \eq{SOLVS3} is
 $\gamma_{\rm SP} = \gamma_{cr} \approx
0.37$. If ${\cal Z} \,\ll \bas \,\,Y$ we can find
 the solution to \eq{SOLVS3} assuming, that
$\gamma_{\rm SP}\,=\,\gamma_{cr} \,+\,\delta \gamma$ with $\delta \gamma\,\ll\,\gamma_{cr}$. The value of $\delta \gamma$ from \eq{SOLVS3} turns out to be equal to
\beq \label{SOLVS4}
\delta \gamma\,\,=\,\,\frac{3}{4} \frac{1}{\Lb 1 - \gamma_{cr}\Rb
 \,\frac{d \chi\Lb \gamma\Rb}{ d \gamma}\Big{|}_{\gamma =
 \gamma_{cr}}} \,\frac{\cal Z}{\bas\,Y}
\eeq

Substituting $\gamma_{\rm SP}\,=\,\gamma_{cr} \,+\,\delta \gamma$ 
 into \eq{SOLVS1} we obtain the expression for the saturation momentum:
\beq \label{SOLVS5}
\xi_s\,=\,\ln\Lb Q^2_s\Rb\,\,=\,\,\bas \frac{\chi\Lb \gamma_{cr}\Rb}{ 1 - \gamma_{cr}} \,Y \,\,-\,\frac{3}{4\Lb 1 - \gamma_{cr}\Rb} {\cal Z}
\eeq

In the vicinity of the saturation scale the scattering amplitude
 shows  geometric scaling behaviour\cite{GS1,LETU,GS2,GS3} and
 has the following form \cite{MUT} for $\tau \sim 1$:
\beq \label{SOLVS6}
N\Lb r, b,Y\Rb\,\,=\,\,N_0 \Lb r^2 \,Q^2_s\Lb Y, b \Rb\Rb^{1 - \gamma_{cr}}\,=\,N_0 \tau^{1 - \gamma_{cr}}
\eeq
where $N_0$ is a constant and we assumed that $\delta \gamma 
\,\ln \tau \,\ll\,1$.

Using \eq{SOLVS5} for $Q_s$ we see that $N\Lb r, b,Y\Rb$ of 
\eq{SOLVS6} can be  re-written in the form:
\beq \label{SOLVS7}
N\Lb r, b,Y\Rb\,\,=\,\,N_0 \Lb r^2 \,Q^2_s\Lb Y, b=0 \Rb\Rb^{1 - \gamma_{cr}}\,e^{ - \frac{3}{4} \Lb \frac{b^4}{4 {\alpha'}^2_{\rm eff}\,Y}\Rb^{1/3}}
\eeq
for the kinematic region
\beq \label{SOLVS8}
 r^2 \,Q^2_s\Lb Y=0\Rb e^{\bas \frac{\chi\Lb \gamma_{cr}\Rb}{ 1 -
 \gamma_{cr}} \,Y \,\,-\,\frac{3}{4\Lb 1 - \gamma_{cr}\Rb} {\cal Z}}\,\,\sim\,\,1
 \eeq
        
     
\subsection{The scattering amplitude  in the deep  saturation region}

 We use the approach of Ref.\cite{LETU} for linearizing \eq{NEEQ} deep
 inside  the saturation region ($r^2\,Q^2_s\Lb Y, b\Rb \,\gg\,1$): we
 assume that $N\Lb r, b,Y\Rb\,\to\,1$ 
 and neglect the contributions, which are proportional to $\Delta^2\Lb 
 r, b,Y\Rb\,\ll\,1$ where $ N\Lb r, b,Y\Rb\,=\,1 \,-\,\Delta\Lb  r, b,Y\Rb 
 $. \eq{NEEQ} reduces to the following linear equation for $\Delta\Lb 
 r, b,Y\Rb$
 \beq \label{DSR1}
 \frac{\partial}{\partial Y} \Delta\Lb \vec{r}, \vec{b} ,  Y \Rb \,\,=\,\,\,- \bas\!\! \int^{r}_{1/Q_s\Lb Y,b\Rb}\frac{d^2 \vec{r}'}{2\,\pi}\,K\Lb \vec{r}', \vec{r} - \vec{r}'; \vec{r}\Rb\,\Delta\Lb \vec{r},\vec{b},Y \Rb\,\,+\,\,\,\Lb \alpha'_{\rm eff}\,\nabla^2_b\Rb^2 \,\Delta\Lb \vec{r},\vec{b},Y \Rb
\eeq
     
     In the previous section we found, that the scattering amplitude
 for $\tau \sim 1$ 
  shows  geometric scaling behaviour, being  a function of  only  one
 variable  $\tau$.      
    The first term in the r.h.s. of \eq{DSR1}   is  equal to $-\bas
 \ln \tau \equiv - \bas \zeta$.  We can re-write
 $\frac{\partial}{\partial Y}$ as 
    \beq \label{DSR2}   
    \frac{\partial}{\partial Y}\,=\,   \frac{\partial}{\partial \zeta}  \frac{\partial \zeta}{\partial Y} \,=\, \Lb\bas \frac{\chi\Lb \gamma_{cr}\Rb}{ 1 - \gamma_{cr}}\,\, +\,\frac{1}{4\Lb 1 - \gamma_{cr}\Rb\,Y} {\cal Z}\Rb\frac{\partial}{\partial \zeta}    \eeq
    Assuming that $ \frac{1}{4\Lb 1 - \gamma_{cr}\Rb\,Y} {\cal Z}\,\,\ll\,\,1$ we can neglect this contribution and re-write \eq{DSR1} in the form
    
        \beq \label{DSR3}
     \bas \frac{\chi\Lb \gamma_{cr}\Rb}{ 1 - \gamma_{cr}} \frac{\partial \Delta\Lb \zeta, b\Rb}{\partial \zeta}  \,=\,-\bas \zeta \,\Delta\Lb \zeta, b\Rb\,\,+\,\,\,  \Lb \alpha'_{\rm eff}\,\nabla^2_b\Rb^2 \,\Delta\Lb \zeta,b \Rb  
\eeq
  where
    \beq \label{DSR4}
    \zeta\,\,=\,\,\xi + \xi_s \,\,\,=\,\xi \,\,+\,\,\bas \frac{\chi\Lb \gamma_{cr}\Rb}{ 1 - \gamma_{cr}}\,Y\,\,\,\,-\,\frac{3}{4\Lb 1 - \gamma_{cr}\Rb} {\cal Z}
    \eeq

The solution to \eq{DSR3} can be written in the form (see section III-A):
\beq \label{DSRSOL}
 \Delta\Lb \zeta, b\Rb\,\,=\,\,\exp\Lb - \frac{\zeta^2}{2 \kappa}\Rb\,\Phi\Lb \zeta, b\Rb
 ~~~\mbox{with} ~~~~~~~\kappa \,=\, \frac{\chi\Lb \gamma_{cr}\Rb}{ 1 - \gamma_{cr}} 
 \eeq    
           
  This solution violates the geometric scaling behaviour of the scattering 
amplitude,
  leading to the additional suppression of the scattering amplitude at large
 $b$. However, it should be emphasized,  that the main dependence on the
 impact parameter is included in $b$ dependence of the variable $\zeta$ (see
  \eq{DSR4}). 
  For the BK equation the asymptotic behaviour of the scattering
 amplitude  is determined by $ \Delta\Lb \zeta, b\Rb\,\,=\,\,{\rm
 Const}\exp\Lb - \frac{\zeta^2}{2 \kappa}\Rb$, with a constant
 which we cannot estimate. The solution of \eq{DSRSOL} states that
    instead of     ${\rm Const}$ we can have function $\Phi\Lb Y,b\Rb$.
 However, in the next section we will argue that actually we can replace
 $\Phi\Lb Y,b\Rb$ by 1.          

\subsection{Instructive example: solution to leading twist non-linear equation.}

       The general non-linear evolution that is given by \eq{NEEQ}
 is difficult to analyze analytically  for the full BFKL kernel of
 \eq{KERC} or \eq{CHI}. This kernel 
  includes the summation over all twist
 contributions. We would like to start  with a simplified version
 of the kernel in which  we restrict ourselves 
 to the leading twist term only, which has the form 
\bea \label{SIMKER}
\chi\Lb \gamma\Rb\,\,=\,\, \left\{\begin{array}{l}\,\,\,\frac{1}{\gamma}\,\,\,\,\,\,\,\,\,\,\mbox{for}\,\,\,\tau\,=\,r Q_s\,<\,1\,\,\,\,\,\,\mbox{summing} \Lb \ln\Lb1/(r\,\Lambda_{\rm QCD})\Rb\Rb^n;\\ \\
\,\,\,\frac{1}{1 \,-\,\gamma}\,\,\,\,\,\mbox{for}\,\,\,\tau\,=\,r Q_s\,>\,1\,\,\,\,\,\mbox{summing} \Lb \ln\Lb r Q_s\Rb\Rb^n;  \end{array}
\right.
\eea
instead of the full expression of \eq{KERC}.

As  indicated in \eq{SIMKER}  we have two types of logs: $ \Big(\bas
 \ln\Lb r\,\Lambda_{QCD}\Rb\Big)^n$ in the perturbative QCD kinematic
 region where  $r\,Q_s\Lb Y,b\Rb\,\,\equiv\,\,\tau\,\,\ll\,\,1$; and
 $  \Big(\bas \ln\Lb r \,Q_s\Lb Y,b \Rb \Rb\Big)^n$ inside the
 saturation domain ($\tau\,\gg\,1$), where $Q_s\Lb Y, b \Rb$ denotes the
 saturation scale. To sum these logs  it is necessary to modify the 
BFKL kernel
 in different ways in the two kinematic regions, as  shown in 
\eq{SIMKER}.

Inside  the saturation region where $\tau\,\,=\,\,r^2\,Q^2_s
\Lb Y,b\Rb\,\,>\,\,1$ the logs 
   originate from the decay of a large size dipole into one small
 size dipole  and one large size dipole.  However, the size of the
 small dipole is still larger than $1/Q_s$. This observation can be
 translated in the following form of the kernel
\beq \label{K2}
 \int \, \displaystyle{K\Lb \vec{x}_{01};\vec{x}_{02},\vec{x}_{12}\Rb}\,d^2 x_{02} \,\,\rightarrow\,\pi\, \int^{x^2_{01}}_{1/Q^2_s(Y,b)} \frac{ d x^2_{02}}{x_{02}^2}\,\,+\,\,
\pi\, \int^{x^2_{01}}_{1/Q^2_s(Y, b)} \frac{ d |\vec{x}_{01}  -
 \vec{x}_{02}|^2}{|\vec{x}_{01}  - \vec{x}_{02}|^2}
\eeq

Inside the saturation region \eq{NEEQ} takes the form
\beq \label{BK2}
\frac{\partial^2 \widetilde{N}\Lb Y; \xi, \vec{b}\Rb}
{ \partial Y\,\partial \xi}\,\,=\,\, \bas \,\left\{ \Lb 1 
\,\,-\,\frac{\partial \widetilde{N}\Lb Y; \vec{r}, \vec{b}
 \Rb}{\partial  \xi}\Rb \, \widetilde{N}\Lb Y; \xi,
 \vec{b}\Rb\right\}\,\,+\,\,\Lb \alpha'_{\rm eff}\,\nabla_b^2\Rb^2 \frac{\partial \widetilde{N}\Lb Y; \xi, \vec{b}
 \Rb}{\partial  \xi}\eeq
where 
 $\widetilde{N}\Lb Y; \xi, \vec{b}\Rb\,\,=\,\,\int^{
r^2} d r'^2\,N\Lb Y; \vec{r}', \vec{b}\Rb/r'^2\,\,=\,\,\int^{\xi} d \xi' N\Lb Y,\xi',\vec{b}\Rb$ .
 
 The advantage of the simplified kernel of \eq{SIMKER} is 
that,  in the  Double Log Approximation (DLA) for $\tau < 1$,
 it provides a matching with the DGLAP evolution equation\cite{DGLAP}.
 
Introducing
\beq \label{BKLT1}
\frac{ \partial \widetilde{N}\Lb Y; \xi, \vec{b}\Rb  }{\partial \xi} \,\,=\,\,1 - e^{- \phi\Lb Y,\xi,\vec{b}\Rb}
\eeq     
  we can re-write \eq{BK2} in the form:
  \beq \label{BKLT2}
    \frac{\partial \phi\Lb Y,\xi,\vec{b}\Rb}{\partial Y}\,e^{- \phi\Lb Y,\xi,\vec{b}\Rb}\,\,=\,\,  \bas  e^{- \phi\Lb Y,\xi,\vec{b}\Rb}  \widetilde{N}\Lb Y; \xi, \vec{b}\Rb\,\,  +\,\,\Lb \alpha'_{\rm eff}\,\nabla_b^2\Rb^2\, e^{- \phi\Lb Y,\xi,\vec{b}\Rb}    
\eeq
 
 Assuming that $\phi\Lb \zeta,b \Rb = \phi\Lb Y, \xi\Rb\,\,+\,\,\tilde{\phi}\Lb Y, b\Rb$ we can re-write \eq{BKLT2} as follows:
   \beq \label{BKLT3}
    \frac{\partial \phi\Lb Y,\xi,\vec{b}\Rb}{\partial Y}\,\,\,=\,\,  \bas  \,  \widetilde{N}\Lb Y; \xi, \vec{b}\Rb\,\,+\,\,\,\,\underbrace{\Bigg(e^{ \tilde{\phi}\Lb Y,\vec{b}\Rb}  \Lb \alpha'_{\rm eff}\,\nabla_b^2\Rb^2\, e^{-\tilde{ \phi}\Lb Y,\vec{b}\Rb} \Bigg) }_{\mbox{  function of $Y, b$}}
\eeq

 Using \eq{DSR2} and \eq{DSR4} with
 \beq \label{BKLT4}
     \zeta\,\,=\,\,\xi + \xi_s \,\,\,=\,\xi \,\,+\,\,4 \bas\,Y\,\,\,\,-\,\frac{3}{2} {\cal Z}
\eeq 
  We can reduce \eq{BKLT3}, applying
 $\frac{\partial}{\partial \xi}$ to the both sides of  this equation, 
 to the following equation 
 \beq \label{BKLT5}
 \frac{d^2 \phi\Lb \zeta\Rb}{d \zeta^2} \,=\,\frac{1}{4}\Lb 1\,\,-\,\,e^{ - \phi\Lb \zeta\Rb}\Rb
 \eeq
 
 Introducing $\frac{d \phi\Lb \zeta\Rb}{d \zeta} \,\,=\,\, F\Lb \phi\Rb $
  we  can re-write \eq{BKLT5} in the form
 \beq \label{BKLT6}
  \h \frac{ d F^2\Lb \phi\Rb}{d \phi}\,=\,\frac{1}{4}\Lb 1
 - e^{ - \phi}\Rb;~~~~~~~
 F^2\Lb \phi\Rb = \int d \phi \frac{1}{2}\Lb 1 - e^{ - \phi}\Rb \,=\, \h \Lb-1 +  \phi + e^{- \phi} \Rb + C|\Lb Y,b\Rb
 \eeq
 Finally,
 \beq \label{BKLT7}
  \sqrt{2} \int^\phi_{\phi_0} \frac{d \phi'}{\sqrt{ -1 + \phi' + e^{-\phi'} + C\Lb Y,b\Rb}}\,=\,\zeta + {\rm Const}
 \eeq
 Further informtion regarding this equation  can be  found in the book of 
Ref.\cite{MATH}( see 
formula {\bf 4.1.1.}).$\phi_0$ is the value of $\phi $ at $\zeta = 0$.
  One can see that if $\phi_0 \,\ll\,1$ we have $\phi = \phi_0 e^{\h 
\zeta}$ for $\zeta \to 0$ for ${\rm Const} =0$ and the arbitrary
 function $C\Lb Y,b\Rb =0$. Taking $\zeta$ from \eq{BKLT4} one can see
 that $\phi =\phi_0 \Lb r^2 \,Q_s\Lb Y, b \Rb\Rb^{1 - \gamma_{cr}}$
 providing the matching of the solution  at $\tau\,= \,r^2
 \,Q^2_s\Lb Y,b\Rb\, >\,1$ with the solution at $\tau \,\leq 1$\,
 (see section IV-B).
 
 For large $\phi$ \eq{BKLT7} leads to $\phi = {\rm Const} \exp\Lb -
 \zeta^2/8\Rb $,   which is the scattering amplitude of \eq{DSRSOL}
 for the our simplified BFKL kernel. However, we find that $\Phi
 =1$ in this solution. 
  
 Therefore, in this simplified version of the non-linear equation we found,
 that the scattering amplitude in the saturation region has the geometric
 scaling behaviour being a function of one variable $\tau = r^2\,Q^2\Lb Y,
 b\Rb$  with 
 \beq \label{QS}
 Q^2_s\Lb Y, b \Rb\,\,\,=\,\,\,Q^2_s\Lb Y=0\Rb\, \exp\Lb\bas \frac{\chi\Lb
 \gamma_{cr}\Rb}{ 1 - \gamma_{cr}} \,Y \,\,-\,\frac{3}{4\Lb 1 - \gamma_{cr}
\Rb} \Lb \frac{b^4}{4 {\alpha'}^2_{\rm eff}\,Y}\Rb^{1/3}\Rb
 \eeq
  
  Hence we  find that the $b$ dependence is concentrated in the 
dependence
 of the saturation scale, as it has been assumed in the numerious attempts 
to introduce $b$-dependence in the saturation models (see Ref.\cite{RESH,CLP}
 and references therein).


\section{The Froissart disc in QCD.}

 In sections IV-C  and IV-D we showed, that (i) the scattering
 amplitude shows  geometric scaling behaviour being a function
 of one variable $\zeta$(see \eq{DSR4})
 and (ii) $ 1 - N\Lb \zeta\Rb \,=\,\Delta\Lb \zeta\Rb \,\ll\,1$ for $\zeta \,\geq\,1$. The condition $\zeta \,\geq\,1$ means that 
 \beq \label{FD1}
 \bas \frac{\chi\Lb \gamma_{cr}\Rb}{ 1 - \gamma_{cr}}\,Y\,\,+\,\,\xi \,\geq\,\,\frac{3}{4\Lb 1 - \gamma_{cr}\Rb}\Lb \frac{b^4}{4\,{\alpha'}^2_{\rm eff}\,Y}\Rb^{1/3}
 \eeq 
 Therefore, for 
 \beq \label{FD2}
 \frac{b}{\sqrt{\alpha'_{\rm eff} }}\,\,\leq\,\,R_{\rm F}\,\,=\,\,\Bigg(4\,Y\,\Big(4\bas \frac{\chi\Lb \gamma_{cr}\Rb}{ 3}\,Y\,\,+\,\,\frac{4 ( 1 - \gamma_{cr})}{3}\,\xi \Big)^3\Bigg)^{1/4}
 \eeq
 where $R_F$ is the radius of the Froissart disc.
 
\begin{figure}[h]
     \begin{center}
     \includegraphics[width=0.75\textwidth]{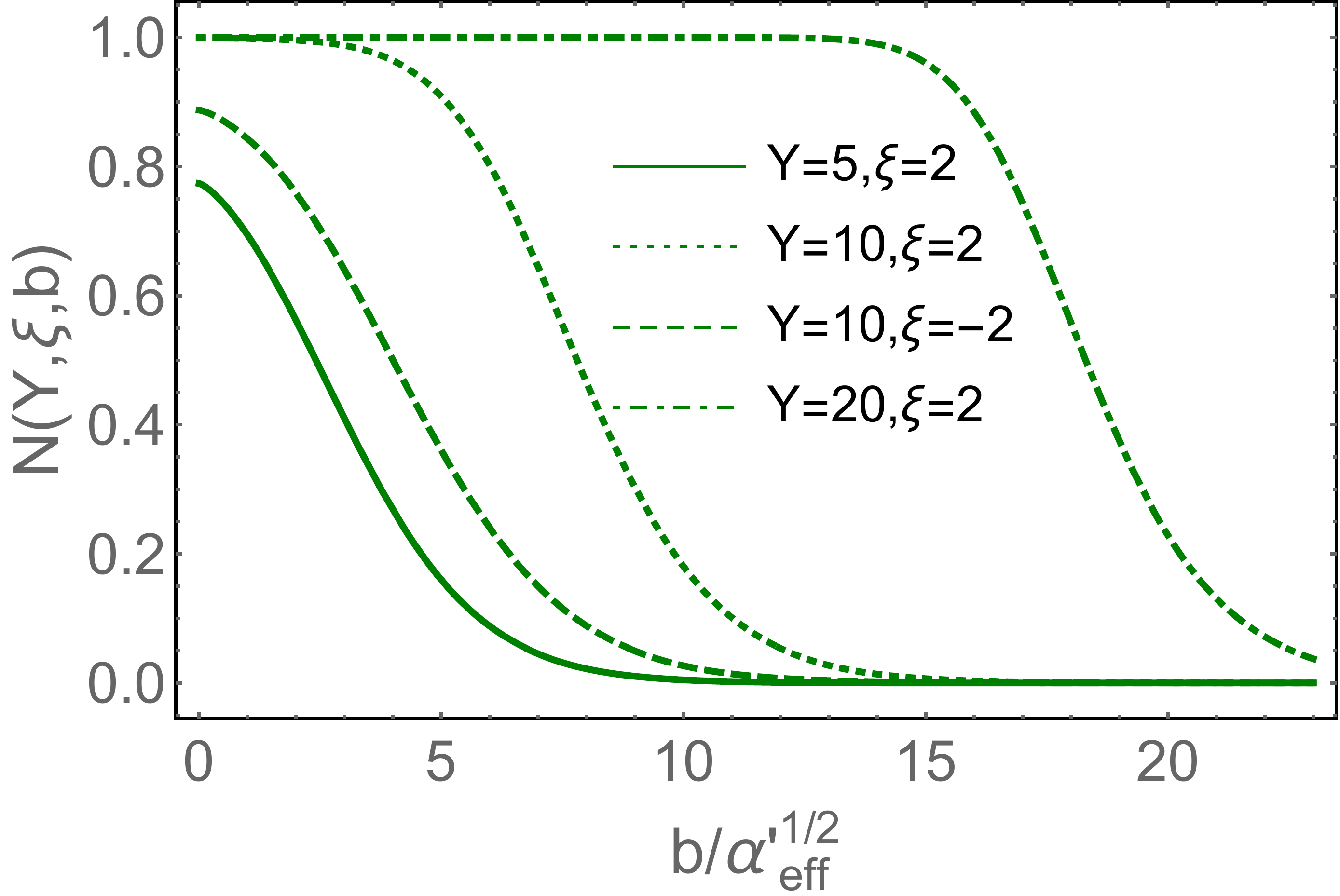} 
     \end{center}    
      \caption{The impact parameter($b$)  dependence  of the scattering
 amplitude. For negative $\zeta$ of \eq{DSR4} we use the solution of
 \eq{SOLVS7}, while for $\zeta \,>\,0$  \eq{BKLT7} is solved with
 $C(Y,b)=0$ and ${\rm Const}=0$. $N_0$ is taken to be equal to
 0.055 and $\bas$ = 0.25.}
\label{bdep}
   \end{figure}

 In \fig{bdep} we plot the solution  to the leading twist non-linear
 equation (see previous section)   as function of $b$.  
   The solution demonstrates all features of the Froissart disc
 with a radius, which is proportional to $Y$. The edge of the disc 
 decreases as $\exp\Lb - C_1\Lb b - R_F\Rb^3\Rb$ where $C_1$ is a
 constant that can be calculated.  Indeed, we can see this plugging
 in the expression  for the amplitude
 $N\Lb r, Y; b\Rb\,\,=\,\,N_0\Lb r^2 \,Q^2_s\Lb Y,b \Rb\Rb^{\bar{\gamma}}$
 $\vec{b}\,=\,\vec{R}_F  + \vec{\Delta b}$ and expanding the amplitude
 with respect to $\Delta b$. One can see, that for $\Delta b\, \ll\,R_F$,
 the amplitude is equal
 \beq \label{FD3}
 N\Lb r, Y; b\Rb\,\,=\,\,N_0\,\exp\Lb - \,\frac{3}{2}\,\Lb \frac{2 R_F}{Y}\Rb^{1/3}\frac{ \Delta b}{\sqrt{\alpha'_{\rm eff}}}\Rb\,\,\xrightarrow{Y\,\gg\,\xi} \,\,\,N_0\,\exp\Lb - 6 \Lb \frac{\chi\Lb \gamma_{cr}\Rb}{3}\Rb^{3/4}\,\frac{ \Delta b}{\sqrt{\alpha'_{\rm eff}}}\Rb \eeq

  For large  $b \,\gg\,R_F$ the typical $ b^2 \propto\,\sqrt{Y}$.

     The new equation leads to restoration of the Froissart theorem.
 Indeed, going back to \eq{FR2}  one can see, that we can use \eq{SOLVS6}
 and \eq{SOLVS7} to estimate  the value of $b_0$. Indeed, this value we
 can find from the following equation:
     
     \beq \label{FD3}
  N\Lb Y, \xi,b_0\Rb\,\,=\,\,\underbrace{e^{\bas \chi\Lb \gamma_{cr}\Rb\,Y \,+\,\Lb 1 - \gamma_{cr}\Rb\,\xi \,-\,\frac{3}{4} \Lb \frac{b_0^4}{4\,{\alpha'}^2_{\rm eff}\,Y}\Rb^{1/3}}}_{\mbox{in vicinity of the saturation scale}}\,\,=\,\,f_0
     \eeq 
   Assuming $Y \,\gg\,\xi$ we obtain 
   
   \beq \label{FD4} 
   \frac{b_0}{\sqrt{{\alpha'}_{\rm eff} }}\,\,=\,\,\Lb\frac{4 \bas \chi\Lb \gamma_{cr}\Rb}{3}\Rb^{1/4}\,\,Y
   \eeq

   Plugging this value for $b_0$ into \eq{FR2} we have
   
   \beq \label{FD5}
   \sigma_{tot}\,\,\leq \,\,2\,\pi\,b^2_0\,=\,2\,\pi\,{\alpha'}_{\rm eff}  \Lb\frac{4 \bas \chi\Lb \gamma_{cr}\Rb}{3}\Rb^{1/2}\,\,Y^2
   \eeq
   and the integral over $b > \,b_0$ gives a smaller contribution,
 which is of the order of $\sqrt{Y}$. \eq{FD5} is the Froissart
 theorem in the specific realization for  the QCD approach, based
 on the new non-linear evolution equation.
   
\begin{figure}[h]
     \begin{center}
     \includegraphics[width=0.65\textwidth]{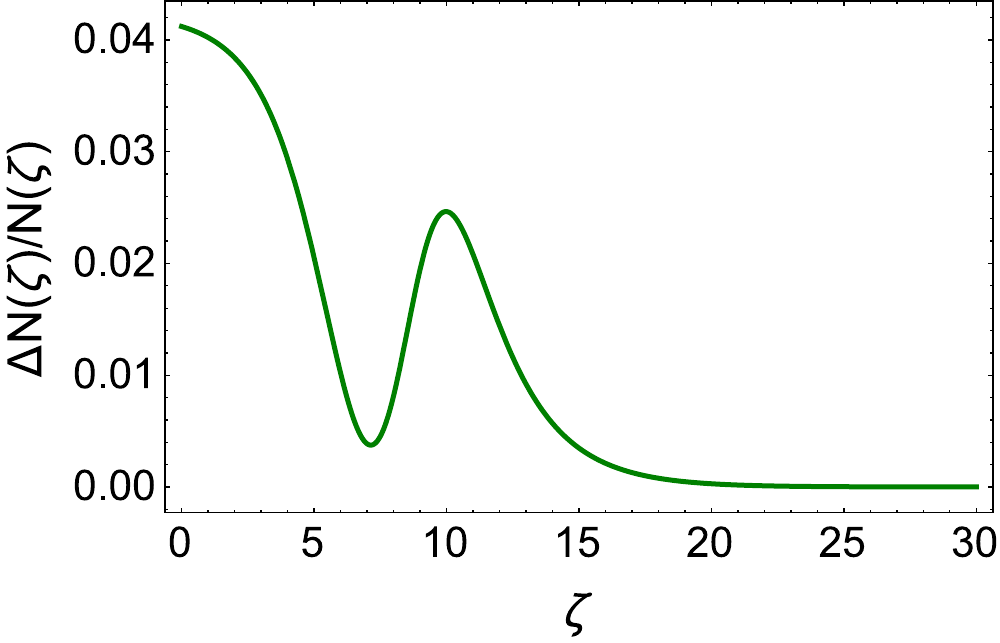} 
     \end{center}    
      \caption{The comparison of the approximate  analytical
 formula of \eq{FD6}   with the numerical solution of \eq{BKLT7}.
 $\Delta N = \Lb N\Lb \zeta, \eq{BKLT7} \Rb \,.-\,
  N\Lb \zeta, \eq{FD6} \Rb\Rb/N\Lb \zeta, \eq{BKLT7} \Rb$.
       \eq{BKLT7} is solved with $C(Y,b)=0$ and ${\rm Const}=0$.
 $N_0$ is taken to be equal to 0.055 and $\bas$ = 0.25. In 
\eq{FD6} $a = 0.65$.}
\label{APF}
   \end{figure}

   For  applications  one can use the simple analytical form of
 the solution to the non-linear equation, given in Ref.\cite{LEPP}:
   \beq \label{FD6} 
   N\Lb \zeta\Rb\,\,=\,\,a\Lb 1 -  \exp\Lb - N_0\,e^{\h\, \zeta}\Rb\Rb\,\,+\,\,(1 - a) \frac{N_0\,e^{\h\, \zeta}}{1\,\,+\,\,N_0\,e^{\h\, \zeta}}
   \eeq
   which describe the numerical solution within the accuracy of
 4\% (see \fig{APF}).

        \section{Conclusions}
   
 The main result of this paper is the new non-linear evolution equation
 which includes the random walk both in the transverse momentum and in
 impact parameter of the produced gluons (dipoles). We showed, that the
 solution to  new equation results in the exponential  decrease of 
the
 scattering amplitude at large impact parameter and in the
 restoration of the Froissart theorem.  Therefore, this equation
 solves the long standing
 problem of the CGC/saturation approach. We demonstrated, that the
 new equation generates the amplitude, which approaches   1
 for $b\, \leq \,{\rm Const} \,Y$ and which  decreases as
 $\exp\Lb - \mu b\Rb$ at $b\, >\, {\rm Const} \,Y $.

 We gave a brief review of the previous attempts to include the
 diffusion in $b$, in the framework of the BFKL equation, and showed
 that the new equation provides a kind of the generalization of all
 these attempts.

 We found the solution to the equation in the kinematic region, 
where ${\cal Z}\,\,\ll\,\,4\,\bar{\gamma}\, Y$.  Hence, a problem
 for the future is to develop a more general solution, as well as to
 describe the available experimental  data within an approach,
 based on the new equation.
 
  {\it Acknowledgements.} \\
   We thank our colleagues at Tel Aviv University and UTFSM for
 encouraging discussions.
 This research was supported  by 
 CONICYT PIA/BASAL FB0821(Chile)  and  Fondecyt (Chile) grant 1180118 .  
 
\end{document}